\documentclass{IEEEoj}
\usepackage{cite}
\usepackage{graphicx, color}
\usepackage{textcomp}
\usepackage{amsmath}
\usepackage{amsfonts}
\usepackage{amssymb}
\usepackage{dsfont}
\usepackage{booktabs}
\usepackage{algorithm}
\usepackage{algpseudocode}
\usepackage[none]{hyphenat}
\usepackage{orcidlink}
\usepackage[printonlyused]{acronym}
\usepackage{tabularx}

\def\BibTeX{{\rm B\kern-.05em{\sc i\kern-.025em b}\kern-.08em
    T\kern-.1667em\lower.7ex\hbox{E}\kern-.125emX}}

\AtBeginDocument{\definecolor{ojcolor}{cmyk}{0.93,0.59,0.15,0.02}}

\def\OJlogo{\vspace{-4pt}\hskip-4pt\includegraphics[height=18pt]{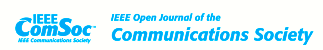}}

\begin{document}

\receiveddate{XX Month, XXXX}
\reviseddate{XX Month, XXXX}
\accepteddate{XX Month, XXXX}
\publisheddate{XX Month, XXXX}
\currentdate{24 July, 2026}
\doiinfo{OJCOMS.2026.011100}

\title{Beam scheduling policy for communications and PNT services from LEO satellites}

\author{
Alejandro Gonzalez-Garrido\IEEEauthorrefmark{1} \IEEEmembership{(Senior Member, IEEE)} \orcidlink{0000-0002-4695-8797}, 
Francesco Menzione \IEEEauthorrefmark{1} \orcidlink{0009-0008-4182-9081}, 
Ottavio M. Picchi\IEEEauthorrefmark{2}, 
and Carla Amatetti \IEEEauthorrefmark{3} \orcidlink{0000-0003-0253-2871}
\IEEEmembership{(Member, IEEE)}
}
\affil{European Commission, Joint Research Centre, Ispra, (Italy)}
\affil{External consultant for European Commission, Joint Research Centre, Ispra, (Italy)}
\affil{University of Bologna, Bologna, (Italy)}
\corresp{CORRESPONDING AUTHOR: Alejandro Gonzalez-Garrido (e-mail: alejandro.gonzalez-garrido@ec.europa.eu).}

\authornote{This work was supported by XXX}
\markboth{Beam scheduling policy for communications and PNT services from LEO satellites}{Alejandro Gonzalez-Garrido \textit{et al.}}

\begin{abstract}
Future low Earth orbit (LEO) constellations are expected to provide positioning, navigation, and timing (PNT) as a native service alongside broadband, removing the GNSS dependency inherited by 5G non-terrestrial networks. This is known in the literature as fused PNT. This paper asks how a multibeam satellites in a LEO constellation should share their beams and power between a communication service (COM) and PNT in the scenario when communications keep full priority. Each beam on a satellite can provide a communication service, a PNT service or both (this under certain limitations). Then, the PNT service is evaluated by the share of area the reach $95\%$-of-time availability for a fixed maximum power available on each satellite. This paper present this PNT availability question as a beam-power budget, as the PNT beams do not need to transmit at the same power as the communication beams thanks to the processing gain on the receiver. Two satellite beam scheduler policies are compared: one where on a cell the beam serving COM demand is exclusive for COM, therefore the PNT service in this cell is provided by the other satellites in line of sight, it requires no signal modification and attains $70.5\%$ / $88.2\%$ availability under homogeneous/population-weighted traffic profiles; and a second policy, the COM beam can provide also PNT signals, this in-beam ranging embeds the ranging signal in the COM waveform, reaching the ceiling at the cost of waveform redesign. The availability gap between the two policies is concentrated on the thin-overlap equatorial belt.
\end{abstract}


\maketitle

\section{Introduction}\label{sec:intro}

\Ac{PNT} has become an invisible utility of modern infrastructure: power-grid synchronization, financial timestamping, transport, and the mobile networks themselves all consume it, overwhelmingly from \ac{GNSS}. Yet \ac{GNSS} signals reach the Earth at power levels on the order of $-155$~dBW from \ac{MEO}, which makes the service structurally exposed to jamming, spoofing, and interference, a vulnerability identified as critical more than two decades ago~\cite{john_a_volpe_national_transportation_systems_center_vulnerability_2001}, and routinely confirmed since \cite{c4ads_above_2019,noauthor_easa_2025}. \Ac{LEO} constellations have emerged as the natural complement, tens of dB less \ac{FSPL}, rapidly changing geometry, and large fleets offer a path to resilient \ac{PNT}~\cite{reid_broadband_2018, prol_position_2022}, whether through dedicated payloads, through opportunistic reuse of communication downlinks~\cite{khalife_first_2022,stock_survey_2024} or fused \ac{PNT} architectures \cite{picchi_fused_2025,m_picchi_fused_2025} where the \ac{PNT} capabilities are embedded within the communication signal in a collaborative manner.

When \ac{3GPP} brought satellites into the \ac{5G} \ac{NR} \ac{NTN} in Release~17, the fast-varying \ac{LEO} channel forced a design shortcut: the \ac{UE} is \emph{assumed to be GNSS-capable}, using its \ac{GNSS}-derived position and the broadcast ephemeris to pre-compensate timing advance and Doppler before it ever talks to the network~\cite{3gpp_38821_2021, figaro_5g_2026}. The communication system therefore depends on an \emph{external} navigation system even to close its own synchronization loops: a \ac{GNSS} outage,  accidental or hostile, degrades not only positioning but connectivity itself, coupling precisely the two services that a \textbf{resilient design should decouple}. \ac{3GPP} has recognized the limitation and is unwinding it release by release: Release~19 introduced first elements of \ac{GNSS}-independent operation and enhanced \ac{NTN} positioning~\cite{3gpp_3gpp_2026, figaro_5g_2026}, and Release~20 studies GNSS-resilient NTN operation under interruption, jamming, and spoofing of the GNSS input~\cite{figaro_5g_2026}.

The \ac{6G} is heading exactly there, the \ac{IMT}-2030 framework from \ac{ITU} elevates high-precision positioning and integrated sensing to native capabilities of the radio network, alongside ubiquitous connectivity explicitly including \ac{NTN} integration~\cite{itu-r_framework_2023}. On the \ac{3GPP} roadmap, the Release~20 \ac{6G} study phase started in late 2025, and the timeline approved in June 2026 makes Release~21 the first normative \ac{6G} release, feeding the \ac{IMT}-2030 submission. With Release~22 continuing the \ac{6G} evolution~\cite{3gpp_timeline_2026}. Within this trajectory, a \ac{6G} \emph{native \ac{PNT} service}, where the network providing position and time as a first-class service rather than assuming them from a third party, are recurring design goals for \ac{6G} \ac{NTN}~\cite{figaro_5g_2026, liao_integration_2023}. Once the constellation itself must produce \ac{PNT} while selling broadband capacity, the central engineering question is no longer \emph{whether} to integrate the two services but \emph{how to share the finite payload resources}, such as active beams and transmit power, between them, slot by slot and cell by cell, under the assumption of needing at least four satellites in \ac{LOS} for positioning ($K_{\min} \ge 4$).

\subsection{State of the art}\label{sec:sota}

\subsubsection{Beam scheduling and resource allocation in broadband constellations}
Dynamic beam resource management is a mature line of research in satellite communications. Beam hopping was introduced for \ac{GEO} high-throughput systems as a means to match a limited number of active beams to a spatially uneven traffic demand, and was shown to outperform conventional static illumination~\cite{anzalchi_beam_2010}. The new-space transition to \ac{LEO} mega-constellations broadened the problem to time-varying geometry and massive fleets, as surveyed in~\cite{kodheli_satellite_2021}. Recent work optimizes the beam illumination pattern jointly with the transmission scheme, e.g.\ beam hopping combined with \ac{NOMA} precoding under per-beam traffic demands~\cite{wang_joint_2022}, or allocates \ac{LEO} beam-hopping resources under spectrum-sharing constraints with \ac{GEO} incumbents~\cite{tang_resource_2021}. Across this literature the optimization objectives are communication-centric (offered-vs-served capacity matching, delay, interference), the tools range from mixed-integer programming to metaheuristics and learning-based policies, and, the payload provides a \emph{single} service: every active beam carries broadband traffic, and performance is measured per beam or per user, never as a multi-satellite geometric condition on the ground cell as for a \ac{PNT} service.

\subsubsection{Positioning from LEO}
The survey in~\cite{prol_position_2022} covers the full design space of \ac{LEO}-\ac{PNT}, from dedicated payloads, hosted signals, to its opportunistic use, together with signal parameters, channel models, and commercial initiatives. The work~\cite{reid_broadband_2018} established the feasibility and expected performance of using broadband \ac{LEO} constellations for navigation, and subsequent experimental work demonstrated opportunistic positioning from non-cooperative signals. The work \cite{stock_survey_2024} is a survey dedicated only to this opportunistic approach to \ac{PNT}. Other works, shows real experimentation results with Starlink or other communication satellites downlinks~\cite{khalife_first_2022, kozhaya_unveiling_2025,jardak_practical_2023, kassas_navigation_2023}. This literature is signal and receiver-centric, the constellation and its transmissions are taken as given, and the resource cost of providing the navigation service on the satellite, like beams, power, and their competition with the communication mission, is not modelled. The geometric prerequisite of positioning is normally assumed as $K_{\min} \ge 4$ simultaneously \ac{LOS} satellites~\cite{kaplan2006understanding}, and some works aim to reduce $K_{\min} \ge 1$ as \cite{wang_location-based_2021,nawaz_rtt_2026}, however, this prerequisite appears as a receiver-side visibility statistic, not as a schedulable design.

\subsubsection{Integrated communication and navigation}
The integration of both functions on the same \ac{LEO} platform has recently been surveyed from the perspective of \ac{6G} \ac{NTN}~\cite{liao_integration_2023} with waveform co-design, clock and orbit-determination architectures, and navigation paradigms that reuse communication infrastructure. Three recent proposals are the closest antecedents of the two policies developed here. The work~\cite{de_gaudenzi_integrated_2024} integrates \ac{PNT} into a beyond-5G \ac{LEO} \ac{NTN} by decoupling the services at payload level: the \ac{PNT} signal is broadcast over wide-area coverage while the communication mission keeps its narrow beams, optimizing each payload chain separately at the price of a dedicated \ac{PNT} aperture. In~\cite{gonzalez-garrido_joint_2025}, conversely, the \emph{same} multibeam payload provides joint communication and navigation, with the ranging function delivered through the communication beam infrastructure itself, the in-beam ranging policy used in this work. Besides, at the signal level, the work \cite{gonzalez-garrido_5g_2026} assesses the impact of transmitting the \ac{5G} \ac{PRS} inside the broadband communication downlink of \ac{LEO}/\ac{MEO} \ac{NTN} constellations, showing that with an appropriate periodic design the positioning signal remains detectable while the broadband service suffers no significant degradation. Providing precisely the co-service coexistence thanks to the power backoff and despreading margins that a cell shared between a \ac{COM} and \ac{PNT} beams of \emph{other} satellites requires. Integration is thus addressed at the architecture and signal levels; the \emph{network} level, deciding slot by slot which cells receive \ac{PNT} beams from which satellites, under a shared beam/power budget and against a live communication demand, remains unformulated: no \ac{ICAN} work poses the allocation with the $K_{\min}$-fold simultaneous-illumination requirement as a constraint, nor evaluates it with the availability metric that navigation practice uses.

\subsection{Gap and contribution}
Three topics approach the problem in the literature from different sides without meeting: beam scheduling optimizes communication alone, LEO-PNT analyses navigation without payload resource scheduling, \ac{ICAN} integrates signals and architectures but nothing about resource allocation at constellation level. The gap is a network-level treatment in which communication and positioning \emph{compete for the same beam and power budget} under the geometric requirement of positioning. This work fills that gap with \emph{scheduler policies} whose optimality is afterwards verified by an optimizer algorithm:

\begin{enumerate}
  \item A network-level system model: every active beam of a Walker constellation~\cite{walker_satellite_1984}, propagated with \ac{SGP4}~\cite{vallado_revisiting_2006}, serves either: broadband demand generated by a reproducible M/Geo/$\infty$ session model, from an homogeneous or population-weighted profile, or a \ac{PNT} broadcast signal requiring $K_{\min} \ge 4$ simultaneous satellites per cell. \ac{PNT} is evaluated as the area that keeps a $95\%$ of time availability during 24h simulation, the same \ac{KPI} used in \ac{GNSS} service standards~\cite{us_department_of_defense_global_2020} evaluations.

  \item The bounds that expose which mechanism set $95\%$ total time availability of \ac{PNT} of a cell.
  \begin{enumerate}
      \item The geometric ceiling is set by the constellation design. 

      \item The \ac{COM} traffic profile effect in the \ac{PNT} distribution.
  \end{enumerate}

  \item Two beam-scheduling policies, both closed form when the priority is always to serve the \ac{COM} cells to maximize the $95\%$ availability of \ac{PNT} service:
    \begin{itemize}
        \item \emph{Co-satellite sharing}: The satellite that provide \ac{COM} service to a cell, do not serve \ac{PNT} in that cell, but the other satellites in view, can provide the ranging signal to this cell. The interference effect of the ranging signal has been demonstrated negligible to the \ac{COM} service in \cite{gonzalez-garrido_5g_2026}.

        \item \emph{In-beam ranging}: The \ac{COM} waveform embeds the ranging signal in the \ac{JCAP} spirit as demonstrated in \cite{gonzalez-garrido_joint_2025}.
    \end{itemize}
    Their trade-off are quantified regionally/globally and per service profile.
\end{enumerate}

Therefore, we model the joint \ac{COM}/\ac{PNT} beam-allocation problem of a \ac{LEO} mega-constellation on an exact equal-area global grid, derive its structural bounds, propose two closed-form scheduler policies (where \ac{COM} is always the priority). The outcome is a design rule rather than an algorithm: with power-differentiated ranging beams, the constellation serves its complete communication demand while the \ac{PNT} availability meets the bound of the chosen coexistence rule, $70.5\%$/$88.2\%$ of the band area under co-satellite sharing (homogeneous/population traffic) and the full $92.9\%$ geometric ceiling under in-beam ranging.

The remainder of the paper is organized as follows: Section~\ref{sec:system_model} develops the system model: grid, constellation, beam and coexistence rules, traffic, and the service metrics; Section~\ref{sec:problem_formulation} derives the performance bounds, presents the two scheduling policies; Section~\ref{sec:results} reports the simulation campaign: scenario statistics, the availability achieved by each policy, and the trade-off between the two solutions; Section~\ref{sec:conclusions} concludes the work and foresee future work.

\section{System Model}\label{sec:system_model}

This section describes the spatial discretization, the constellation and beam geometry, the association between beams and cells, the service and traffic models, and the performance metrics. The allocation of beams to services, which is the subject of the policy problem, is addressed in Section~\ref{sec:problem_formulation}.

We consider a \ac{LEO} constellation whose satellites carry a digitally steered multibeam payload and the payload can provide two services per beam: a \ac{COM} and/or a \ac{PNT} service. Then, the Earth surface is partitioned into a discrete grid of equal-area hexagonal/pentagonal cells, and each satellite beam illuminates exactly one cell. Along the text we use the term beam to refer to the transmission from the satellite to ground and the term cell to the specific area within the Earth grid. Besides, time is divided into slots during which the geometry is considered frozen.

\subsection{Discrete Global Grid} \label{sec:grid}

\begin{figure}[h]
    \centering
    \includegraphics[width=\linewidth]{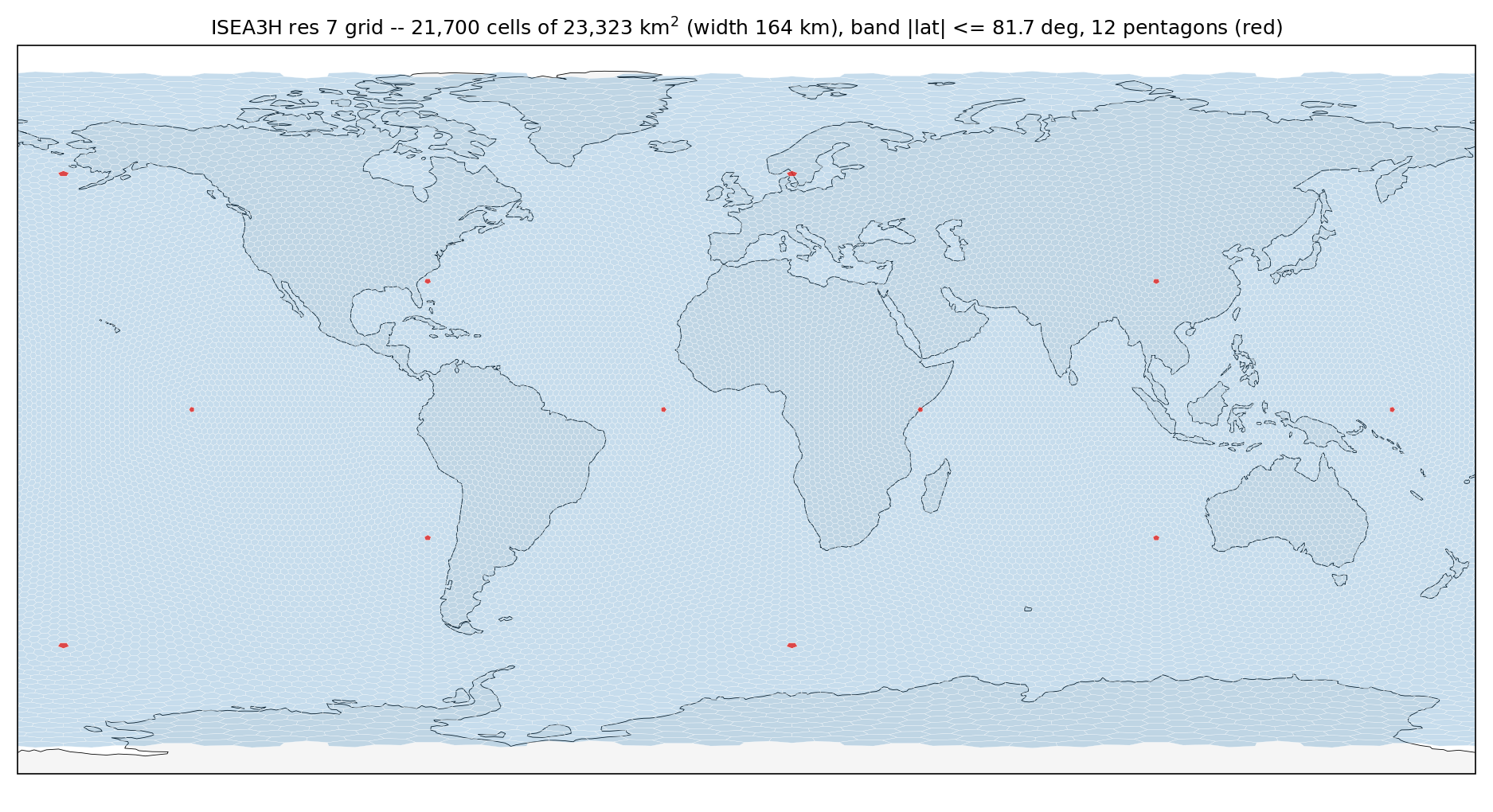}
    \caption{Hexagonal global grid with 150~km diameter cell. In red the 12 pentagons needed.}
    \label{fig:hexagonal_global_grid}
\end{figure}

The constellation service area is discretized with the \ac{ISEA3H}, a discrete global grid system that partitions a sphere into cells of equal area~\cite{sahr_geodesic_2003}. At resolution $r$ the grid contains
\begin{equation}
  M_r = 10 \cdot 3^{r} + 2
  \label{eq:cell_count}
\end{equation}
cells, of which exactly $12$ are pentagons centred at the vertices of the underlying icosahedron and the remaining $M_r - 12$ are hexagons. The grid is defined on the equal area sphere of radius $R$, whose surface area equals that of the reference ellipsoid \ac{WSG84}. Since each pentagon has five sixths of the hexagon area, the equal-area property yields the exact cell areas, as seen in Figure~\ref{fig:hexagonal_global_grid}
\begin{equation}
  A_{\mathrm{hex}}(r) = \frac{4 \pi R^{2}}{10 \cdot 3^{r}},
  \qquad
  A_{\mathrm{pen}}(r) = \frac{5}{6}\, A_{\mathrm{hex}}(r).
  \label{eq:cell_areas}
\end{equation}
A regular hexagon of area $A$ has a width across flats of
\begin{equation}
  w = \sqrt{ 2 A / \sqrt{3} },
  \label{eq:cell_width}
\end{equation}
which is used below as the characteristic cell size.

Each beam is designed to illuminate one cell in an Earth fixed scheme~\cite{3gpp_38821_2021}. Denoting by $D$ the beam footprint diameter across flats, the footprint area is $A_{\mathrm{b}} = (\sqrt{3}/2) D^{2}$, and the grid resolution is selected as the one whose hexagon area is closest to the footprint area on a logarithmic scale,
\begin{equation}
  r^{\star} \;=\; \Big\lfloor \log_{3}\!\Big(
      \frac{4 \pi R^{2}}{10\, A_{\mathrm{b}}} \Big) \Big\rceil ,
  \label{eq:resolution}
\end{equation}
where $\lfloor \cdot \rceil$ denotes rounding to the nearest integer. In the remainder of the paper the resolution index is omitted and the cell set is denoted by $\mathcal{G} = \{1, \dots, M\}$, restricted to the latitude band actually reachable by the constellation, as specified in subsection~\ref{sec:beams}. A cell $g \in \mathcal{G}$ has an area $A_g \in \{A_{\mathrm{hex}}, A_{\mathrm{pen}}\}$ and centroid position $R\,\mathbf{u}_g$, where $\mathbf{u}_g \in \mathbb{R}^{3}$ is the unit vector of the cell centre in an \ac{ECEF} frame.

\begin{figure}
    \centering
    \includegraphics[width=\linewidth]{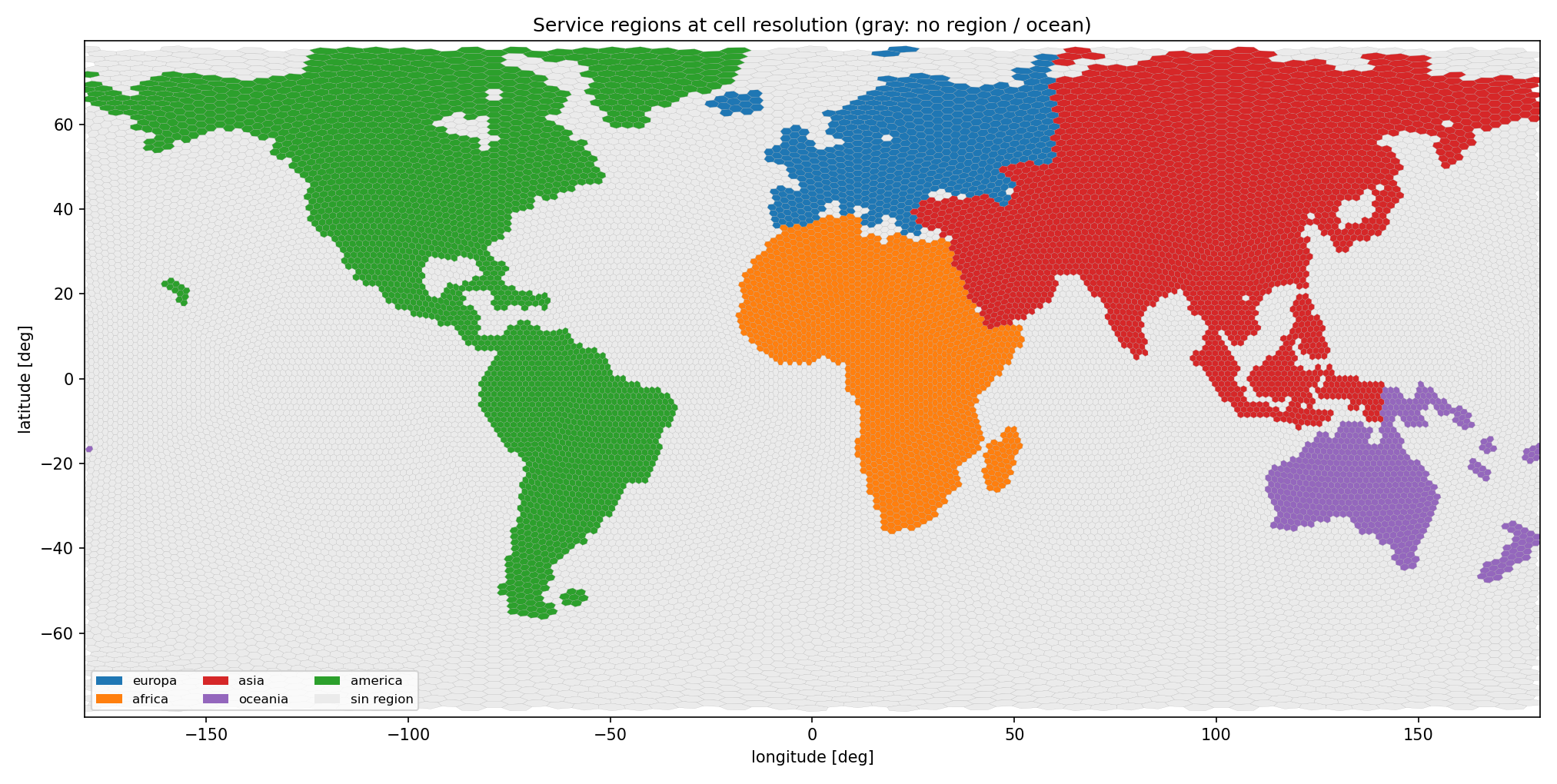}
    \caption{Cells marked for the different regions used for the services priority}
    \label{fig:cells_regions}
\end{figure}

Finally, we also divide the cells $\mathcal{G}$ by regions as seen in Figure \ref{fig:cells_regions} (Europe, Africa, Asia, Oceania and America), this way, over certain areas of interest a service (\ac{COM} or \ac{PNT}) can have different priority. These regions just apply a certain weight $w_g$ to the cell $g$, and this weight is taken into account on the performance metrics for each service.

\subsection{Constellation Geometry}
\label{sec:constellation}

The space segment is a Walker delta constellation $i{:}S/P/F$ with $S = PQ$ satellites in near-circular orbits of altitude $h$ and inclination $i$, arranged in $P$ equally spaced orbital planes with $Q$ satellites per plane and phasing factor $F \in \{0, \dots, P-1\}$~\cite{walker_satellite_1984}. The right ascension of the ascending node of plane $p$ and the initial mean anomaly of satellite $q$ in plane $p$ are
\begin{equation}
  \Omega_p = \Omega_0 + \frac{2\pi p}{P},
  \qquad
  M_{p,q}(t_0) = \frac{2\pi q}{Q} + \frac{2\pi F p}{S},
  \label{eq:walker}
\end{equation}
for $p = 0, \dots, P-1$ and $q = 0, \dots, Q-1$. 

\begin{figure}[h]
    \centering
    \includegraphics[width=\linewidth]{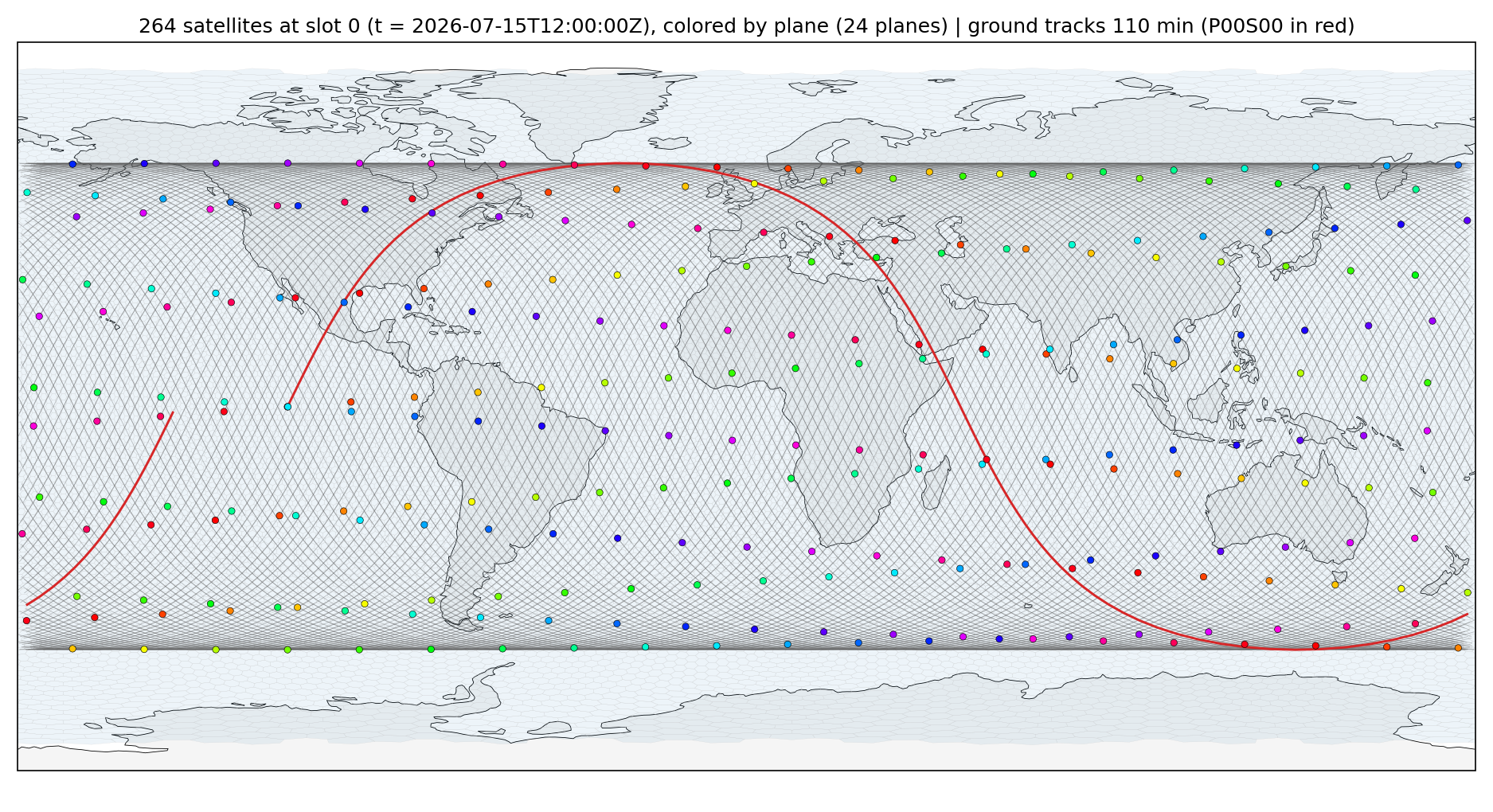}
    \caption{Constellation deployment with the ground track for 110 minutes for a single satellite.}
    \label{fig:satellite_deployment}
\end{figure}

The satellites are indexed by $s \in \mathcal{S} = \{1, \dots, S\}$ and their \ac{ECEF} positions $\mathbf{r}_s(t)$ are obtained by propagating the orbital elements with the \ac{SGP4} model, which accounts for the dominant Earth oblateness perturbation~\cite{vallado_revisiting_2006}. The sub-satellite point of satellite $s$ at time $t$ is the geodetic projection of $\mathbf{r}_s(t)$ onto the Earth surface, with unit vector $\hat{\mathbf{u}}_s(t)$ as seen in Figure \ref{fig:satellite_deployment}.

\subsection{Beam Model and Cell Association}
\label{sec:beams}

Each satellite $s\in S$ generates a cluster of $N_{\mathrm{b}}$ spot beams matched to the grid, arranged as a centred cluster of $n_{\mathrm{r}}$ rings around the nadir beam, so that
\begin{equation}
  N_{\mathrm{b}} = 1 + 3\, n_{\mathrm{r}} (n_{\mathrm{r}} + 1).
  \label{eq:beam_count}
\end{equation}

\begin{figure}[h]
    \centering
    \includegraphics[width=\linewidth]{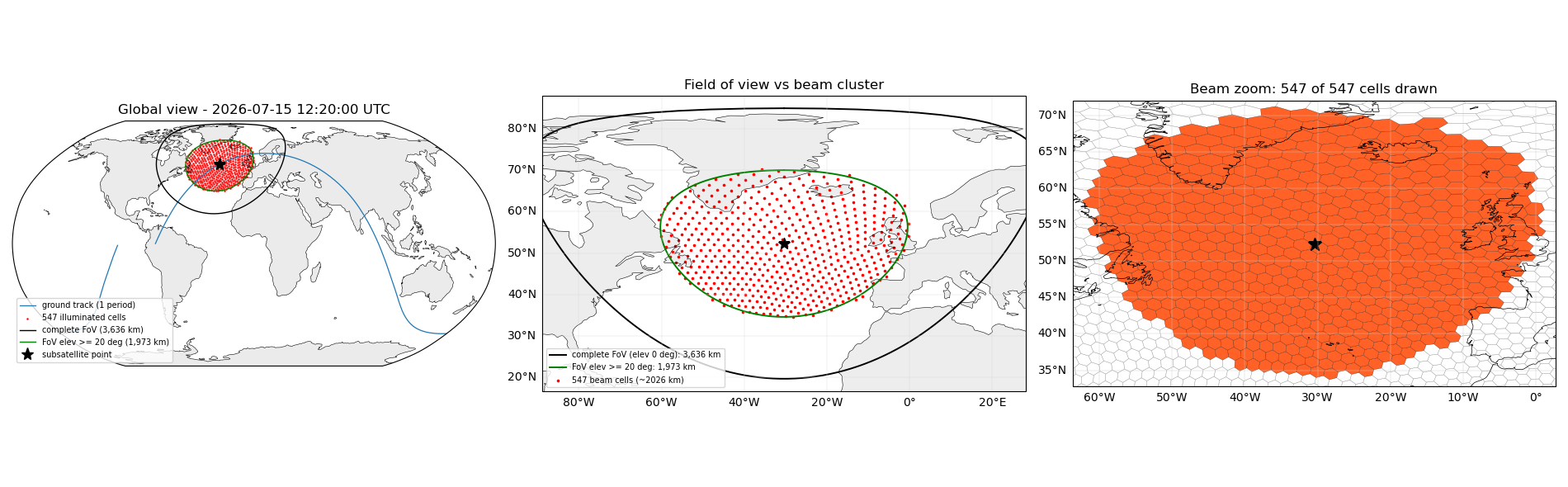}
    \caption{Single satellite: a) global view of a single satellite $s$, b)\ac{FoV} (black) and elevation angle mask (green) areas, c) hexagonal cells coverage $\mathcal{F}_s(t)$}
    \label{fig:single_satellite_cells_and_coverage}
\end{figure}

The beams are Earth fixed at the slot time scale: the payload steers each beam to the centre of one grid cell, so that beams and cells are in one-to-one correspondence within the satellite footprint. The footprint of satellite $s$ at time $t$, denoted $\mathcal{F}_s(t) \subset \mathcal{G}$, is the set of the $N_{\mathrm{b}}$ cells closest to the sub-satellite point in great-circle distance,
\begin{equation}
  \mathcal{F}_s(t) = \operatorname*{arg\,max}_{
      \mathcal{A} \subset \mathcal{G},\, |\mathcal{A}| = N_{\mathrm{b}}}
      \; \sum_{g \in \mathcal{A}} \mathbf{u}_g^{\mathsf T}
      \hat{\mathbf{u}}_s(t),
  \label{eq:footprint}
\end{equation}
which forms a approximately circular cluster of hexagonal/pentagonal cells centred at nadir as depicted in Figure \ref{fig:single_satellite_cells_and_coverage} right. Since the cells have equal area, the footprint covers a spherical cap of area $N_{\mathrm{b}} A_{\mathrm{hex}}$ whose angular radius seen from the Earth centre is
\begin{equation}
  \psi = \arccos\!\Big( 1 - \frac{N_{\mathrm{b}} A_{\mathrm{hex}}}{2 \pi R^{2}} \Big).
  \label{eq:cap_radius}
\end{equation}

The elevation angle at which the edge of the footprint sees the satellite is
\begin{equation}
  \varepsilon(\psi) = \arctan\!\left( \frac{\cos\psi - R/(R+h)}{\sin\psi} \right),
  \label{eq:edge_elevation}
\end{equation}
so the pair $(N_{\mathrm{b}}, D)$ is chosen such that $\varepsilon(\psi)$ remains above the minimum elevation mask of the user terminals. Figure~\ref{fig:single_satellite_cells_and_coverage} shows an example of a single satellite over the hexagonal grid with its \ac{FoV}, the elevation angle mask at $20^\circ$ and the hexagonal cells covered.

Consistently, the cell set $\mathcal{G}$ is restricted to the latitude band $|\phi_g| \le i + \psi$ plus a small guard band, outside of which no coverage is possible (no satellite \ac{FoV} reach this area). The coverage state of the satellite $s$ at time $t$ is captured by the binary visibility indicators
\begin{equation}
  v_{s,g}(t) = \mathds{1} \{ g \in \mathcal{F}_s(t) \},
  \label{eq:visibility}
\end{equation}
where $\mathds{1} \{ \}$ is the indicator function.

From $v_{s,g}(t)$ we define, for each cell $g$, the  set of satellites in view and its multiplicity,
\begin{equation}
  \mathcal{V}_g(t) = \{ s \in \mathcal{S} : v_{s,g}(t) = 1 \},
  \quad
  m_g(t) = |\mathcal{V}_g(t)|.
  \label{eq:multiplicity}
\end{equation}

The slant range between satellite $s$ and cell $g$ is $\rho_{s,g}(t) = \lVert \mathbf{r}_s(t) - R\,\mathbf{u}_g \rVert$, and the serving \ac{COM} satellite of cell $g$ is defined as the closest visible one,
\begin{equation}
  s^{\star}(g,t) = \operatorname*{arg\,min}_{s \in \mathcal{V}_g(t)}\rho_{s,g}(t),
  \label{eq:owner}
\end{equation}
where, in case of two satellites at the same distance, is selected the lowest satellite index. Selecting the closest satellite maximizes the elevation angle and minimizes the free-space path loss, and therefore corresponds to the standard cell association rule of terrestrial and non-terrestrial networks.

\subsection{Time Discretization}\label{sec:time}

The total time for simulation $T$ is divided into $K$ slots of duration $\Delta t = T/K$, indexed by $t \in \mathcal{T} = \{1, \dots, K\}$, and all geometric quantities are evaluated at the slot start and held constant within the slot. The slot duration $\Delta t$ is matched to the Earth's grid granularity: the ground-track speed of a satellite is approximately $v_{\mathrm{g}} = 2 \pi R / T_{\mathrm{orb}}$, with $T_{\mathrm{orb}}$ the orbital period, and $\Delta t$ is chosen on the order of
\begin{equation}
  \Delta t  \lessapprox \frac{w}{v_{\mathrm{g}}},
  \label{eq:slot_duration}
\end{equation}
so that satellite footprints $\mathcal{F}_s(t)$ advance by about one cell per slot $t$ and the discrete sequence $\{\mathcal{F}_s(t)\}_t$ does not skip cells.

\subsection{COM Traffic Model} \label{sec:traffic}

The \ac{COM} service is modelled as the aggregate traffic of all devices of a cell. A cell with active demand is served when its serving satellite $s^{\star}(g,t)$ allocates one beam in \ac{COM} mode to the cell $g$ on ground during the slot $t$. Only the closest satellite to the demanding cell may carry the \ac{COM} traffic of this cell, in accordance with the association rule in \eqref{eq:owner}.

The traffic demand of each cell is generated by a session-level traffic model with temporal persistence. Sessions arrive at cell $g$ according to a Poisson process with rate $\lambda_g$ sessions per slot, and each ongoing session terminates at the end of every slot independently with probability $1/\bar{L}$, surviving with probability $1-1/\bar{L}$. Session durations are therefore geometrically distributed with mean $\bar{L}$ slots, and the number of active sessions $N_g(t)$ evolves to $N_g(t+1)$ as the discrete-time infinite-server (M/Geo/$\infty$) queue
\begin{equation}
  N_g(t+1) = \mathrm{Bin}\!\left( N_g(t),\, 1 - 1/\bar{L} \right)+\mathrm{Poi}(\lambda_g),
  \label{eq:traffic_chain}
\end{equation}
independently across cells. The first term is a \emph{binomial thinning} of the session population: since every one of the $N_g(t)$ sessions in progress tosses its own independent coin and survives the slot with probability $1-1/\bar{L}$, the number of survivors is by definition a binomial random variable with $N_g(t)$ trials and success probability $1-1/\bar{L}$. This per-session independent pruning is exactly equivalent to drawing a geometric duration for each session at its arrival: by the memoryless property of the geometric law, the probability that a session ends in the current slot never depends on how long it has already been active, so the aggregate one-slot effect on the population is the binomial above. The second term, $\mathrm{Poi}(\lambda_g)$, adds the sessions that arrived during the slot. The demand indicator seen by the scheduler is $\delta_g(t) = \mathds{1}\{ N_g(t) > 0 \}$.

The chain is initialized in its stationary regime, in which $N_g(t)$ is Poisson distributed with mean equal to the offered load $a_g = \lambda_g \bar{L}$ Erlang, so that the stationary demand probability of cell $g$ is
\begin{equation}
  \Pr\{ \delta_g(t) = 1 \} = 1 - e^{-a_g}.
  \label{eq:busy_prob}
\end{equation}
This produces demand that persists over consecutive slots, with mean busy periods governed by $\bar{L}$, while remaining fully reproducible from the rate profile $\{\lambda_g\}$, the mean duration $\bar{L}$ and a random seed. 

\begin{figure}[h]
    \centering
    \includegraphics[width=\linewidth]{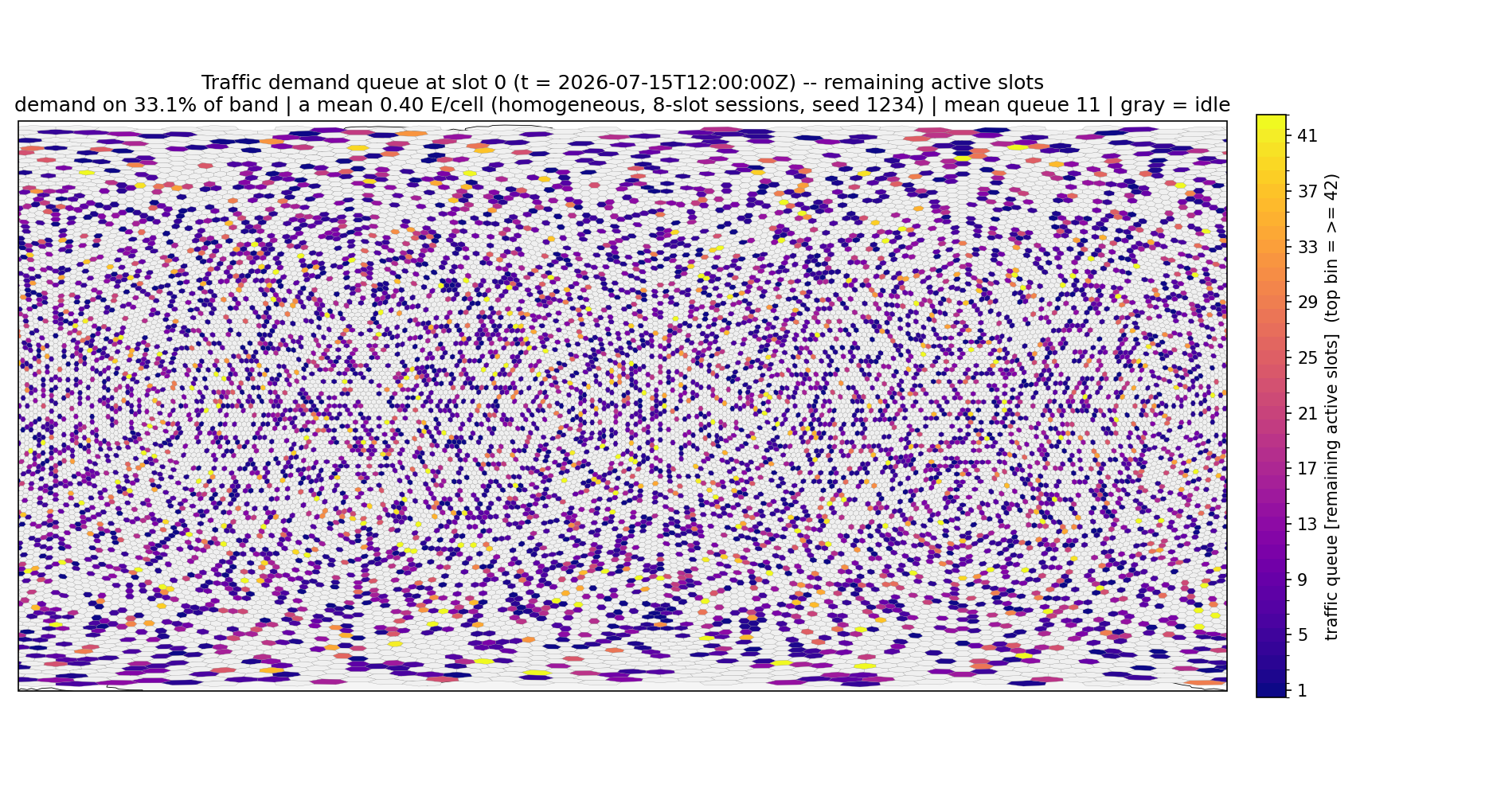}
    \caption{Homogeneous profile realization snapshot example for traffic demand at slot $t=0$.}
    \label{fig:traffic_homogeneous}
\end{figure}

Two spatial profiles of the arrival rates are considered in this work; both share the per-cell dynamics \eqref{eq:traffic_chain} and, by construction, the same total offered load $\sum_g a_g = M\bar{a}$, so their policy results are directly comparable:

\paragraph{Homogeneous profile.}
Every cell offers the same load: $\lambda_g = \lambda$ and $a_g = \bar{a} = \lambda\bar{L}$ for all $g$. Traffic demand is then spatially uniformly distributed, at any instant a fraction $1-e^{-\bar{a}}$ of the cells are busy, with no geographic structure. An example of this profile is shown in Figure~\ref{fig:traffic_homogeneous}.

\begin{figure}[h]
    \centering
    \includegraphics[width=\linewidth]{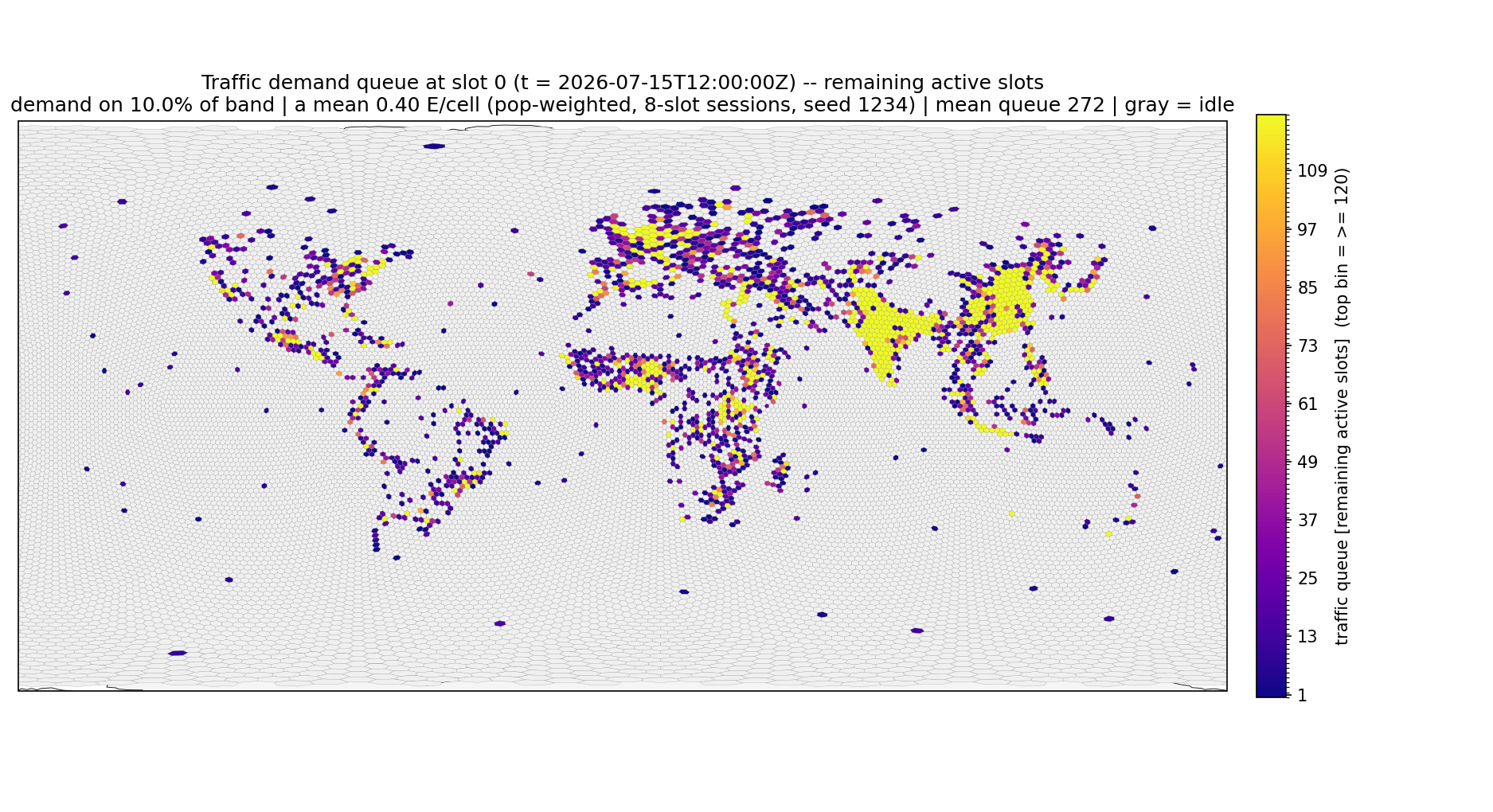}
    \caption{Snapshot of traffic demand at slot $t=0$ for a population-density profile.}
    \label{fig:traffic_pop}
\end{figure}

\paragraph{Population-weighted profile.}
The \ac{COM} traffic demand follows the population density of the Copernicus \ac{GHSL} raster (GHS-WUP, epoch 2025; persons per 1-km pixel on the equal-area Mollweide grid~\cite{melchiorri_ghs-wup-country-stats_2025}). The raster is block-summed to coarse equal-area pixels and each pixel's population is assigned to the grid cell whose centroid is nearest on the sphere, a mass-conserving zonal aggregation that yields the cell populations $P_g$ and shares $s_g = P_g / \sum_h P_h$. The \ac{COM} traffic demand is then distributed proportionally to population, subject to a floor that keeps every cell, including oceanic and unpopulated ones, where sporadic users (ships, aircraft, remote terminals) must still be served, at a small but non-zero rate:
\begin{equation}
  a_g \;=\; \max\!\left( \varphi\,\bar{a},\; \beta\, s_g \right),
  \quad
  \lambda_g \;=\; a_g / \bar{L},
  \label{eq:pop_load}
\end{equation}
with floor fraction $\varphi \ll 1$ (here $\varphi = 10^{-2}$) and $\beta$ the constant that renormalizes the total budget to $\sum_g a_g = M\bar{a}$ exactly: cells pinned at the floor are fixed and the remaining budget is redistributed proportionally to the shares of the unpinned cells (a monotone water-filling that terminates in a finite number of steps). The resulting mean demand probability, $\tfrac{1}{M}\sum_g (1-e^{-a_g})$, is markedly lower than in the homogeneous case for the same total load, because the load concentrates on few, heavily saturated cells (areas like India or China), the spatial concentration the scheduler must then resolve. An example of this profile is shown in Figure~\ref{fig:traffic_pop} for a slot time $t$.

\subsubsection{Performance metric}
The communications service is quantified over the complete simulation time $T$ by the served \ac{COM} demand and its normalized throughput,
\begin{equation}
    \begin{split}
        J_{\mathrm{COM}} &= \sum_{t \in \mathcal{T}} \sum_{g \in \mathcal{G}} w_g c_g(t),\\
        \eta_{\mathrm{com}} &= \frac{J_{\mathrm{COM}}} {\sum_{t} \sum_{g} \delta_g(t)\, \mathds{1}\{ m_g(t) \ge 1 \}},
    \end{split}
    \label{eq:objectives}
\end{equation}
where $J_{\mathrm{COM}}$ counts served cell-slots of demand, weighted by region $w_g$, and $\eta_{\mathrm{com}} \in [0,1]$ is the fraction of the cells that meet the requirements to be served with respect to the cells that are actually served. Ideally all \ac{COM} traffic demand has to be served by the system $\eta_{\mathrm{com}} \rightarrow 1$. The weights $w_g$ are used to give priority over certain areas of interest by the network operator.

\subsection{PNT model}

The \ac{PNT} service is a one-way ranging broadcast, where satellites transmit synchronized ranging signals, and a receiver estimates its position and its clock bias from the pseudoranges, which requires $K_{\min} \geq 4$ simultaneous transmitters, as in \ac{GNSS}~\cite{kaplan2006understanding}. Signals from different satellites share the cell resource through code division, as is standard in satellite navigation (\ac{5G} \ac{PRS} is a form of code division, as each sequence is generated by a different ID)~\cite{gonzalez-garrido_5g_2026}. 

We adopt an ``at least $K_{\min}$'' service rule: cell $g$ is declared \ac{PNT} served at slot $t$ when $K_{\min}$ distinct satellites of $\mathcal{V}_g(t)$ can allocate a \ac{PNT} beam to the cell. An optimal allocation never spends more than $K_{\min}$ ranging signals, not all $m_g$ visible ones, since additional beams consume beams budget. We assume a \ac{LOS} scenario, therefore the user on ground does not need any extra beam. A future work should evaluate and optimize the use of more beams than $K_{\min}$ to cope with urban scenarios, where some blockage is expected, and improved \ac{GDOP}.

\subsubsection{Performance Metrics}

The \ac{PNT} service is evaluated by its \emph{availability}, what matters to a positioning user in cell $g$ is not the instantaneous extent of the \ac{PNT} footprint, but the fraction of time the service can be relied upon at that location. Let
\begin{equation}
  \alpha_g = \frac{1}{K} \sum_{t \in \mathcal{T}} \pi_g(t)
  \label{eq:pnt_availability}
\end{equation}
be the per-cell \ac{PNT} availability, the fraction of the $K$ slots in which cell $g$ enjoys valid \ac{PNT} service. The system-level metric is the share (regional/global) of the service area that meets a target minimum $95\%$ time availability level $\gamma$,
\begin{equation}
  \eta_{\mathrm{pnt}}^{\gamma} =
  \frac{\displaystyle \sum_{g \in \mathcal{G}} A_g w_g \mathds{1}\{ \alpha_g \ge \gamma \}}
       {\displaystyle \sum_{g \in \mathcal{G}} A_g w_g},
  \quad \gamma = 0.95,
  \label{eq:normalized_metrics}
\end{equation}
where the weights $w_g$ represent the priority of the cell $g$. Then, the fraction of the total covered area whose cells keep valid \ac{PNT} during at least $\gamma=95\%$ of the temporal slots. This value $95\%$ is the standard availability level of navigation service specifications~\cite{us_department_of_defense_global_2020}. 

The exact cell areas in \eqref{eq:cell_areas}, including the twelve pentagons, make the area share exact. Three properties of this metric are worth noting: 
\begin{enumerate}
    \item It is bounded by the constellation design, independent of any scheduling decision: valid \ac{PNT} requires $m_g(t) \ge K_{\min}$ satellites in view, so $\alpha_g \le \tilde{\alpha}_g = \frac{1}{K}\sum_t \mathds{1}\{ m_g(t) \ge K_{\min} \}$, and therefore $\eta_{\mathrm{pnt}}^{\gamma}$ can never exceed the area share with $\tilde{\alpha}_g \ge \gamma$.
    
    \item It is stricter than the time-averaged served area $\frac{1}{K}\sum_t \sum_g A_g\, \pi_g(t)$ that the per-slot allocation distributes: a schedule that sweeps a large but rotating footprint may attain a high mean served area while leaving every individual cell below the $\gamma$ threshold, whereas $\eta_{\mathrm{pnt}}^{\gamma}$ only credits cells that are revisited persistently, at least a fraction $\gamma$ of the slots.
    
    \item It can have a regional view: the weights $w_g$ let the operator prioritize the availability of selected regions. Setting $w_g > 1$ on a region increases its share in \eqref{eq:normalized_metrics} and, since the same weights drive any scheduling decision under scarcity, steers the service toward it; uniform weights $w_g = 1$ recover the plain area share, and the per-region evaluation of $\eta_{\mathrm{pnt}}^{\gamma}$ (restricting numerator and denominator to one region) is used in Section~\ref{sec:results} to localize where each solution gains or loses. Unless stated otherwise, the results use uniform weights.
\end{enumerate}

\section{Beam scheduling policy design and optimal validation}\label{sec:problem_formulation}

The beam allocation problem inherits the structure of the system model: all geometric quantities are frozen within a slot, the traffic indicator $\delta_g(t)$ is known at scheduling time, and no constraint couples different time slots, the beam power budget is a per-slot power limit, not an energy reservoir. The section proceeds in two steps. First, closed-form bounds locate what any scheduler could at best achieve and which mechanism, geometry or beam-power budget, binds. Second, two scheduling \emph{policies} are proposed for the operating point of practical interest, communications with full priority: both turn out to be closed form, no per-slot optimization is required to run them.

\subsection{PNT availability metric analysis}

The availability metric for \ac{PNT} in \eqref{eq:normalized_metrics} is bounded by two independent mechanisms, and identifying which one binds tells the designer where capacity can actually be bought:
\begin{equation}
  \eta_{\mathrm{pnt}}^{\gamma} \le \min\!\left( \eta_{\mathrm{geo}}^{\gamma},\; \eta_{\mathrm{bud}}^{\gamma} \right).
  \label{eq:eta_bounds}
\end{equation}

The \emph{geometric ceiling} $\eta_{\mathrm{geo}}^{\gamma}$ is the area share of cells whose eligibility availability satisfies $\tilde{\alpha}_g = \frac{1}{K}\sum_t \mathds{1}\{ m_g(t) \ge K_{\min} \} \ge \gamma$, in other words, no schedule can serve a cell in a slot in which fewer than $K_{\min}$ satellites see it. The \emph{budget bound} $\eta_{\mathrm{bud}}^{\gamma}$ follows from the beam budget $N_{\mathrm{max}}$, a compliant cell only needs service in $\lceil \gamma K \rceil$ of the $K$ slots, i.e.\ it may \emph{rest} in up to a $(1-\gamma)$ fraction, and rotating the resting beams subset across the compliant set stretches the budget by $1/\gamma$. Charging every compliant cell its minimum consumption against the aggregate fleet budget $S N_{\max} K$ beam-slots gives
\begin{equation}
  \sum_{g \in \mathcal{P}} \gamma\, \bar{b}_g \le S N_{\max},
  \label{eq:budget_rotation}
\end{equation}
with $\bar{b}_g$ the mean beam cost over the cell's cheapest $\lceil \gamma K \rceil$ eligible slots, filling \eqref{eq:budget_rotation} with the cheapest eligible cells gives $\eta_{\mathrm{bud}}^{\gamma}$, and a cyclic rest rotation attains it exactly when $\gamma K$ is an integer.

For the reference instance of this work (Walker $24\times11$, $S = 264$, $i = 55^\circ$, $h = 1200$~km, $N_{\max} = 150$, $K_{\min} = 4$, $M = 21{,}380$ cells, $K = 1440$ slots of $60$~s, a full day) the two bounds are far apart. Geometry is generous, $\eta_{\mathrm{geo}}^{0.95} = 92.9\%$ of the total coverage area ($19{,}865$ cells). The budget is not. 

If the policy uses all satellites in view for \ac{PNT}, a valid \ac{PNT} cell requires every satellite of the visible set $\mathcal{V}_g$ to point one beam at cell $g$. Therefore, the beam cost of a \ac{PNT} cell is $b_g = m_g(t) \in [4, 14]$, and the budget sustains only $7{,}845$ cells (mean cost $5.31$ beams per served slot), i.e.
\begin{equation}
  \eta_{\mathrm{bud}}^{0.95} = 36.7\% \;\ll\; \eta_{\mathrm{geo}}^{0.95}= 92.9\%,
\end{equation}
a factor $2.5$ below the ceiling. Moreover, the affordable cells are confined to the equatorial belt (median $|{\rm lat}| = 10.5^\circ$), where the overlap $m_g$, and hence the cost, is lowest. The $\pm45$--$55^\circ$ latitude bands, although the best served geometrically ($m_g \approx 10$--$14$), are the most expensive to hold if the policy is to use all satellites in view. Below is the alternative proposed in this scenario.

\paragraph{Relaxing the service rule to ``at least $K_{\min}$''.}
Requiring the full visible set is unnecessarily conservative: a valid fix only needs $K_{\min}$ ranging signals, not $m_g$. Therefore we have defined $\pi_g(t) = 1$ if at least $K_{\min}$ of the $m_g(t)$ visible satellites serve $g$ makes the cost uniform, $b_g = K_{\min} = 4$, so the budget sustains $\lfloor S N_{\max} / K_{\min} \rfloor = 9{,}900$ cells persistently and, with the rest rotation of \eqref{eq:budget_rotation}, $\lfloor S N_{\max} K / (K_{\min} \lceil \gamma K \rceil) \rfloor = 10{,}421$ compliant cells:
\begin{equation}
  \eta_{\mathrm{bud}}^{0.95}:\; 36.7\% \;\longrightarrow\; 48.7\%.
  \label{eq:availability_k_min}
\end{equation}
The uniform cost also decouples \emph{where} to serve from the overlap geometry, a $\pm50^\circ$ cell drops from $10$--$14$ beams to $4$.

\paragraph{Alternatives to increase the \ac{PNT} availability.}
The \ac{PNT} availability metric still quite low as seen in \eqref{eq:availability_k_min}. Table~\ref{tab:levers} quantifies the options on the scenario to improve this metric, all evaluated on top of the ``at least $K_{\min}$'' rule with the rotation of the rest of the beams included.
\begin{itemize}
    \item Increasing $N_{\max}$ buys availability linearly until the ceiling is met at $N_{\max} = 286$.

    \item Aided positioning (a barometric altimeter or a stable receiver clock removes one unknown, $K_{\min} = 3$) cuts the cost of every cell by one fourth and, as a side effect, raises the geometric ceiling itself to $95.4\%$ (cells with $m_g = 3$ become eligible).

    \item Wider \ac{PNT} beams enter quadratically through the cell count ($M \propto d^{-2}$), reaching the ceiling at $d = 207$~km.

    \item Lowering the availability target $\gamma$ is already captured by the $1/\gamma$ rotation gain, which grows into genuine time-sharing at $\gamma \le 1/2$.

    \item The metric's denominator is itself a requirement choice: restricted to the inhabited area ($6{,}948$ cells, $32.5\%$ of the band, holding $100\%$ of the population), the $10{,}421$ sustainable cells cover it entirely and $\eta_{\mathrm{pnt}}^{0.95}$ over inhabited area reaches its own ceiling of $96.0\%$ (at $i = 55^\circ$ the highest-latitude settlements sit at the edge of the reliable band), with $3{,}473$ spare cells ($16.2\%$ of the band) left for oceanic corridors.
\end{itemize}

Excluded here by the requirement of a common \ac{COM}/\ac{PNT} beam lattice, but conceptually the limit of this parameter is the GNSS-like architecture in which ranging is broadcast on an much wider beam and do not consume spot beams altogether as in~\cite{gonzalez-garrido_joint_2025, de_gaudenzi_integrated_2024}.

\begin{table}[h]
    \centering
    \caption{Parameters to tune for the $95\%$-availability bound \eqref{eq:eta_bounds}. Reference instance, ``at least $K_{\min}$'' rule with rest rotation, full budget devoted to \ac{PNT}. Ceiling, due to constellation design, at $92.9\%$.}
    \label{tab:levers}
    \small
    \begin{tabularx}{\columnwidth}{@{}p{0.28\columnwidth} X c@{}}
        \toprule
        Parameter & Setting & $\min(\eta_{\mathrm{geo}}, \eta_{\mathrm{bud}})$ \\
        \midrule
        (baseline)
        & $N_{\max}=150$, $b_g=4$
        & $48.7\%$ \\

        Beam budget $N_{\max}$
        & $200$ / $250$ / $286$
        & $65.0$ / $81.2$ / $92.9\%$ \\

        Aided fix, $K_{\min}=3$
        & altimeter or clock aiding
        & $65.0\%$ \\

        \ac{PNT} beam width $d$
        & $200$~km / $207$~km
        & $86.7$ / $92.9\%$ \\

        Target level $\gamma$
        & $0.50$ / $0.33$ (time-sharing)
        & $92.6$ / $95.2\%$ \\

        Inhabited-area metric
        & denominator $=$ populated cells
        & $96.0\%$ \\
        \bottomrule
    \end{tabularx}
\end{table}

\subsection{Power-differentiated ranging beams and the loaded satellite} \label{sec:power_diff}

The final alternative to improve the \ac{PNT} availability is the one selected in this work. With $K_{\min}$ and the beam lattice fixed by the system requirements, the remaining parameter is the power budget behind $N_{\max}$. The equal per-beam cost implicitly assumes \ac{PNT} and \ac{COM} beams are equally expensive, but they could not be: the ranging signal is a waveform whose receiver recovers the pseudorange measurements through correlation, so the processing gain allows the \ac{PNT} spot beams to be transmitted at a fraction of the \ac{COM} power while preserving the minimum required post-correlation $C/N_0$~\cite{gonzalez-garrido_5g_2026}. Let $X \ge 1$ denote the COM-to-\ac{PNT} per-beam power ratio: a \ac{PNT} beam is transmitted at $1/X$ of the \ac{COM} beam power. The per-satellite power constraint becomes
\begin{equation}
  \sum_{g} x_{s,g}(t) + \frac{1}{X} \sum_{g} y_{s,g}(t) \le N_{\max},
  \label{eq:weighted_budget}
\end{equation}
kept in exact integer arithmetic as $X \sum x + \sum y \le X N_{\max}$ with $X \in \mathbb{N}$. Holding the ceiling set requires $K_{\min} \lceil \gamma K \rceil \times 19{,}865$ \ac{PNT} beam-slots against the $S N_{\max} K$ power-unit-slots of the fleet, so the \ac{PNT} availability ceiling is reached with
\begin{equation}
  X \;\ge\; \frac{\gamma\, K_{\min}\, M_{\mathrm{geo}}^{\gamma}}{S N_{\max}}
  \;=\; \frac{0.95 \times 79{,}460}{39{,}600} \;=\; 1.91,
  \label{eq:c_threshold}
\end{equation}
i.e.\ two \ac{PNT} beams per COM-beam power, a mere $-2.8$~dB of per-beam power backoff. Reserving budget for the \ac{COM} service tightens the threshold only mildly (Table~\ref{tab:cpnt}): serving the mean serviceable demand of the homogeneous traffic profile ($\approx 6{,}880$ cells per slot at unit cost) requires $X \ge 2.31$ ($-3.6$~dB), and of the population-weighted profile ($\approx 2{,}170$ cells) $X \ge 2.02$ ($-3.0$~dB). These thresholds are not so severe: with typical spreading gains, link-budget values of $X$ fall in the $10$--$100$ range, an order of magnitude above \eqref{eq:c_threshold}.

\begin{table}[h]
    \centering
    \caption{Minimum \ac{PNT}-to-COM power ratio $X$ (and per-beam backoff) to reach the $92.9\%$ \ac{PNT} availability  ceiling, as a function of the \ac{COM} load simultaneously served.}
    \label{tab:cpnt}
    \small
    \begin{tabularx}{\columnwidth}{@{}X c c@{}}
        \toprule
        \ac{COM} served concurrently & $X$ & per-beam power loss \\
        \midrule
        None (pure-PNT bound)
          & $1.91$ & $-2.8$~dB \\

        Homogeneous traffic ($\approx 6{,}880$ cells/slot)
          & $2.31$ & $-3.6$~dB \\

        Population-weighted ($\approx 2{,}170$ cells/slot)
          & $2.02$ & $-3.0$~dB \\
        \bottomrule
    \end{tabularx}
\end{table}

\begin{figure*}[t]
  \centering
  \includegraphics[width=0.49\textwidth]{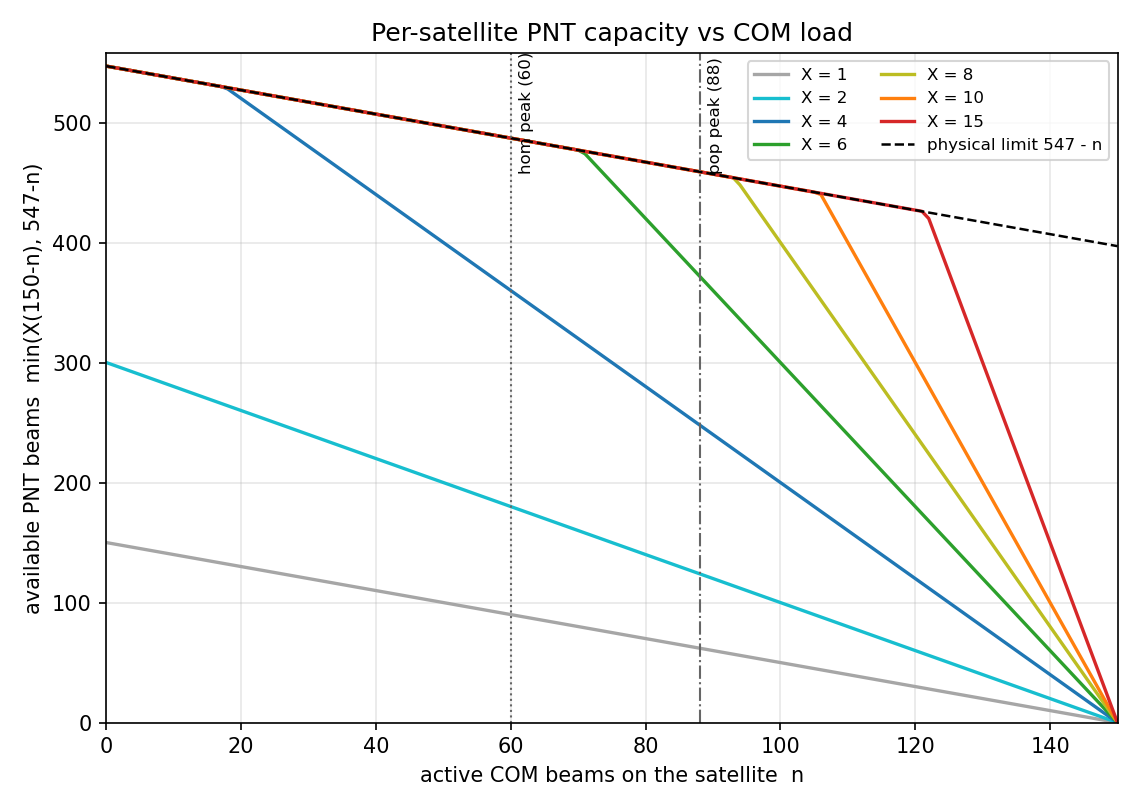}\hfill
  \includegraphics[width=0.49\textwidth]{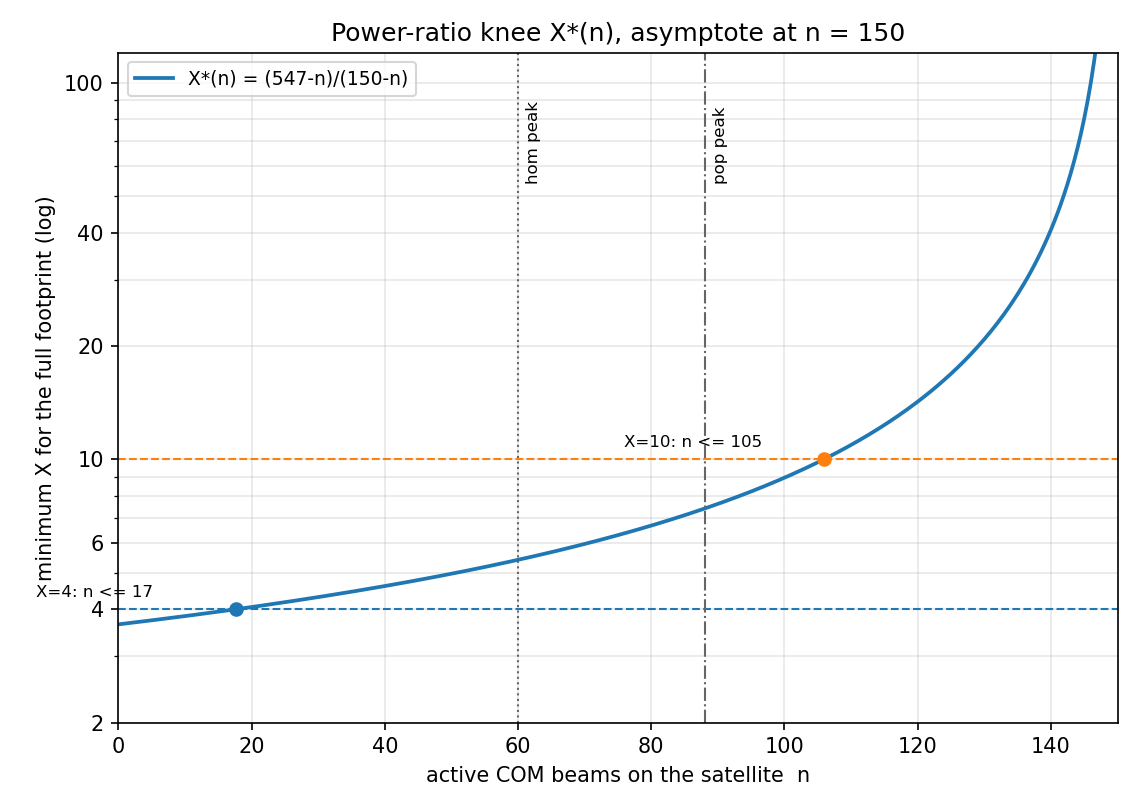}
  \caption{Per-satellite \ac{PNT} capacity versus \ac{COM} load. Left: available \ac{PNT} beams \eqref{eq:pnt_capacity} for several power ratios $X$, against the physical envelope $N_b - n$; the vertical lines mark the worst measured per-satellite \ac{COM} loads of the two traffic profiles. Right: the full-footprint threshold $X^{*}(n)$ of \eqref{eq:xstar} (log scale) with its vertical asymptote at $n = N_{\max}$; the markers show the maximum load $n_{\max}(X)$ tolerated by $X = 4$ and $X = 10$.}
  \label{fig:xload}
\end{figure*}

The aggregate thresholds of Table~\ref{tab:cpnt} hide the per-satellite structure: \ac{COM} demand is spatially concentrated, so the binding question is what a \emph{loaded} satellite can still do for \ac{PNT}. With $n$ \ac{COM} beams active, a satellite retains $X(N_{\max} - n)$ power units and $N_b - n$ physical beams, so its \ac{PNT} capacity is
\begin{equation}
  \mathrm{PNT}(n, X) \;=\; \min\!\bigl( X (N_{\max} - n),\; N_b - n \bigr),
  \label{eq:pnt_capacity}
\end{equation}
with two regimes separated by the full-footprint threshold
\begin{equation}
  X^{*}(n) \;=\; \frac{N_b - n}{N_{\max} - n},
  \qquad
  n_{\max}(X) \;=\; \frac{N_{\max} X - N_b}{X - 1},
  \label{eq:xstar}
\end{equation}
plotted in Fig.~\ref{fig:xload}.

Below $X^{*}$ the satellite is \emph{power-limited}: capacity grows linearly with $X$ and every admitted \ac{COM} beam displaces $X$ \ac{PNT} beams. Above it, the satellite is \emph{physically limited} by its $N_b = 547$ beams: extra power buys nothing and a \ac{COM} beam costs exactly one beam. The threshold starts at $X^{*}(0) = N_b / N_{\max} = 3.65$ (an unloaded satellite lights its whole footprint from $X = 4$, i.e.\ $-6$~dB) and diverges as $n \to N_{\max}$, with sharply diminishing returns: the marginal load tolerance is $dn_{\max}/dX = (N_b - N_{\max})/(X-1)^2$, about $45$ beams per unit of $X$ at $X=4$ but only $5$ at $X = 10$. Note also that $X = 2$ never leaves the power-limited regime ($2 N_{\max} = 300 < N_b$), which is why it reaches the fleet-aggregate ceiling of Table~\ref{tab:cpnt}, where cells only need $K_{\min}$ beams spread over their overlap, but cannot light full footprints. On the scenario, the measured per-satellite \ac{COM} load at full demand service has mean $26.1$ / $8.2$ beams and worst satellite-slot peaks of $60$ / $88$ beams for the homogeneous / population-weighted profiles respectively. The population profile carries less total load but hits harder peaks over megacity passes. By \eqref{eq:xstar}, $X = 6$ and $X = 8$ cover those peaks exactly, and $X = 10$ ($-10$~dB, comfortably above typical processing gains) tolerates up to $105$ concurrent \ac{COM} beams, a $19\%$ margin over the worst observed load.

\subsection{Two scheduling policies} \label{sec:solutions}

\subsubsection{Mode Exclusivity}

The first beam policy \ac{COM} and \ac{PNT}, cannot share a satellite beam: the beam that satellite $s$ points at cell $g$ during a slot carries either the high-power \ac{COM} signal or the low-power ranging waveform, not both. The exclusion, however, is a \emph{per-beam} constraint, not a per-cell one. If the serving satellite $s^{\star}(g,t)$ transmits \ac{COM} on cell $g$, only its own beam is occupied; the remaining satellites of the visible set $\mathcal{V}_g(t)$ may still transmit \ac{PNT} on that cell, because the ranging signal is a waveform radiated at a fraction of the \ac{COM} power: after despreading (assuming a ranging estimation like in \ac{GNSS}\cite{kaplan2006understanding}), the ranging receiver operates below the \ac{COM} carrier, and conversely the low power spectral density of the \ac{PNT} beams leaves the \ac{COM} link essentially unaffected~\cite{gonzalez-garrido_5g_2026}. A cell carrying \ac{COM} can therefore still obtain a valid \ac{PNT} fix, provided $K_{\min}$ ranging beams arrive from satellites \emph{other} than the \ac{COM} server, which effectively requires $m_g(t) \ge K_{\min} + 1$ on that cell. At most $K_{\min}$ ranging beams are admitted on a \ac{COM}-served cell,  additional ones bring no positioning benefit and only add interference or consume satellite resources, so the aggregate interference seen by the \ac{COM} receiver is bounded by construction. The coexistence is made robust by a \emph{two-tier power discipline}: ranging beams aimed at a \ac{COM}-served cell back off to a fraction $1/X_{\mathrm{hot}}$ of the \ac{COM} beam power (e.g.\ $X_{\mathrm{hot}} = 20$, $-13$~dB), so that the aggregate carrier-to-interference floor seen by the \ac{COM} receiver, $10\log_{10}(X_{\mathrm{hot}}/K_{\min})$~dB before the spreading discrimination, remains compatible with the \ac{COM} link budget, while beams aimed at traffic-free cells keep the nominal, higher ranging power $1/X_{\mathrm{free}}$. The dimensioning condition of $X_{\mathrm{hot}}$ falls on the ranging receiver, which faces the near--far problem of recovering a signal $10\log_{10}(X_{\mathrm{hot}})$~dB below the co-channel \ac{COM} carrier: the despreading gain must exceed that backoff by at least the post-correlation signal-to-noise ratio the tracking loops require, a condition comfortably met by \ac{GNSS}-class spreading factors. Interference between beams pointing at different cells is neglected, which is justified by the matched beam and cell layout and by the spatial isolation of the spot beams. An alternative in which the \ac{COM} waveform itself embeds the ranging signal, so the serving beam counts as one of the $K_{\min}$ sources and a busy cell needs no extra satellite, is analyzed as the second solution of this work (Section~\ref{sec:solutions}). As a conservative baseline we also evaluate the legacy cell-blanking rule, in which a \ac{COM} transmission excludes every \ac{PNT} beam on the cell; Section~\ref{sec:results} shows that under sustained traffic this variant collapses the \ac{PNT} availability of precisely the inhabited cells.

Combining the rules above, the realized service state of the network at slot $t$ is described by two binary indicators per cell, no longer mutually exclusive: $c_g(t) = 1$ if the traffic demand of cell $g$ is served in \ac{COM} mode, which requires $m_g(t) \ge 1$ and one \ac{COM} beam from $s^{\star}(g,t)$; and $\pi_g(t) = 1$ if cell $g$ is \ac{PNT} served, which requires one \ac{PNT} beam from at least $K_{\min}$ distinct satellites of $\mathcal{V}_g(t)$, none of which may be the \ac{COM} server when $c_g(t) = 1$. The possible states of a cell are:

\begin{enumerate}
    \item Not covered by any satellite beam. This depends on the constellation design: cells with $|\phi_g| > i + \psi$ are permanently out of reach, and cells near the band edge may be momentarily uncovered.

    \item Idle, $c_g(t) = \pi_g(t) = 0$ without demand. There is coverage from at least one satellite, $m_g(t) \ge 1$, but the users of the cell do not demand traffic and no \ac{PNT} service is being provided.

    \item With traffic demand but unserved, $c_g(t) = 0$. There is coverage, $m_g(t) \ge 1$, and the users of the cell request traffic, but the satellites do not have enough available beams to serve it.

    \item \ac{COM} served only, $c_g(t) = 1$, $\pi_g(t) = 0$. The serving satellite $s^{\star}(g,t)$ dedicates one beam to the traffic of the cell, and fewer than $K_{\min}$ of the remaining visible satellites transmit \ac{PNT} on it.

    \item \ac{PNT} served only, $\pi_g(t) = 1$, $c_g(t) = 0$. At least $K_{\min}$ of the $m_g(t) \ge K_{\min}$ visible satellites dedicate a ranging beam to the cell, and no \ac{COM} traffic is being served on it.

    \item \ac{COM} and \ac{PNT} served, $c_g(t) = \pi_g(t) = 1$. The dual state enabled by the per-beam exclusion: $s^{\star}(g,t)$ carries the traffic of the cell while $K_{\min}$ of the other visible satellites provide the ranging service, which requires $m_g(t) \ge K_{\min} + 1$.
\end{enumerate}

Finally, the satellite's payload can keep at most $N_{\max} \le N_{\mathrm{b}}$ beams simultaneously active in a slot, which models the power budget of the on-board amplifiers. The ratio $N_{\max}/N_{\mathrm{b}}$ is the beam duty factor of the payload.

\subsubsection{Co-satellite sharing with adaptive \ac{PNT} power.}
Equations \eqref{eq:pnt_capacity}--\eqref{eq:xstar} suggest an operating scheme that removes the scheduling problem altogether in the nominal regime. Instead of a fixed design constant, let each satellite divide its leftover power evenly among its remaining beams: with $n$ \ac{COM} beams active, every \ac{PNT} beam is transmitted at a fraction $1/X(n)$ of the \ac{COM} power, operating exactly at the full-footprint threshold of \eqref{eq:xstar} until the receiver-driven limit is hit,
\begin{equation}
  X(n) \;=\; \min\bigl(\, X^{*}(n),\; X_{\max} \,\bigr),
  \label{eq:c_adaptive}
\end{equation}
which is $-5.6$~dB for an unloaded satellite ($X^{*}(0) = 3.65$) and sinks, as the \ac{COM} load grows, toward $X_{\max}$, where the deepest per-beam backoff for which the post-correlation $C/N_0$ still supports acquisition and tracking. 

While every satellite satisfies $n_s(t) \le n_{\max}(X_{\max})$, the trivial policy \emph{serve everything}, all serviceable \ac{COM} demand plus every admissible ranging beam is feasible and, notably, optimal: it maximizes both objectives simultaneously, the \ac{COM} service is complete, and the \ac{PNT} availability meets the eligibility ceiling of the coexistence rule, so no scheduler can improve on it. In this regime the joint \ac{COM}/\ac{PNT} problem is solved by power adaptation alone. In our scenario, setting $X_{\max} = 10$ the limit sits at $105$ beams, above the worst measured peak of $88$, therefore, the nominal system is self-sufficient at the design point.

\paragraph{Two-tier ranging power.}
The interference analysis of the dual \ac{COM}--\ac{PNT} state splits the ranging power into two classes: \ac{PNT} beams aimed at a \ac{COM}-served cell back off to $1/X_{\mathrm{hot}}$ of the \ac{COM} power to protect the carrier, while beams on traffic-free cells keep the nominal $1/X_{\mathrm{free}}$. Because the quietest beams are also the cheapest, the full-footprint condition of \eqref{eq:xstar} relaxes to
\begin{equation}
  n \;+\; \frac{h}{X_{\mathrm{hot}}} \;+\; \frac{N_b - n - h}{X_{\mathrm{free}}} \;\le\; N_{\max},
  \label{eq:two_tier}
\end{equation}
where $h$ is the number of \ac{COM}-served cells of \emph{other} satellites inside the footprint, a large population under load, with measured peaks of $215$ (homogeneous) and $272$ (population-weighted) cells. On the scenario with $X_{\mathrm{hot}} = 20$, condition \eqref{eq:two_tier} holds at \emph{every} satellite and slot already at $X_{\mathrm{free}} = 5$ (homogeneous) and $X_{\mathrm{free}} = 6$ (population-weighted). The quiet-cell beams may transmit at $-7.0$ / $-7.8$~dB, louder than the $-7.8$ / $-9.0$~dB the single-tier scheme requires to cover the same measured peaks, while the dual-cell beams sit at $-13$~dB where the interference budget demands it. The two-tier scheme therefore dominates the single-tier one simultaneously in interference into the \ac{COM} link, in ranging $C/N_0$ on quiet cells, and in power-budget margin.

\subsubsection{In-beam ranging.}
There is a residual gap of the previous analysis, and it is structural, not a scheduling loss as a \ac{COM}-served cell needs $K_{\min}$ ranging beams from the \emph{other} visible satellites. Therefore, the cells that cross their busy slots with $m_g = K_{\min}$ exactly exhaust the $(1-\gamma)$ miss budget by geometry alone, and no beam allocation can recover them. 

The improvement proposed now removes the $+1$ requirement at its source, the ranging signal is embedded within the \ac{COM} waveform itself, as validated in~\cite{gonzalez-garrido_joint_2025}. The serving beam is simultaneously the $K_{\min}$-th ranging source and a \ac{COM}-served cell needs only $K_{\min}-1$ beams from the $m_g - 1$ remaining satellites. The eligibility condition returns to $m_g \ge K_{\min}$, identical to a traffic-free cell. The \ac{PNT} availability geometric ceiling then becomes reachable, again by the serve-all policy with no scheduling: power feasibility is strictly easier than under the previous proposal (a dual served cell (\ac{COM} and \ac{PNT}) consumes one \ac{PNT} beam fewer), and the achieved availability equals the ceiling on both traffic profiles. The price for this alternative moves from the network to the signal: the \ac{COM} waveform must carry embedded ranging (pilots or a spread ranging component underneath the data channel), the payload must keep the two components phase-coherent with the system timescale, and the \ac{PNT} receiver must extract pseudoranges from a communication signal instead of a dedicated ranging waveform. Since achieving the geometric ceiling attains a universal upper bound of \eqref{eq:eta_bounds}, the optimality of this policy needs no solver.

\section{Simulation and Results}\label{sec:results}

\subsection{Implementation} \label{sec:implementation} 

The complete toolchain, campaign generation, traffic replay, policy evaluation, and the availability evaluation, is deterministic given the constellation parameters, the traffic seed, and $(X_{\mathrm{free}}, X_{\mathrm{hot}})$, so every figure of this section is reproducible from a single configuration line.

\subsection{Scenario and coverage statistics}

\begin{figure}
    \centering
    \includegraphics[width=\linewidth]{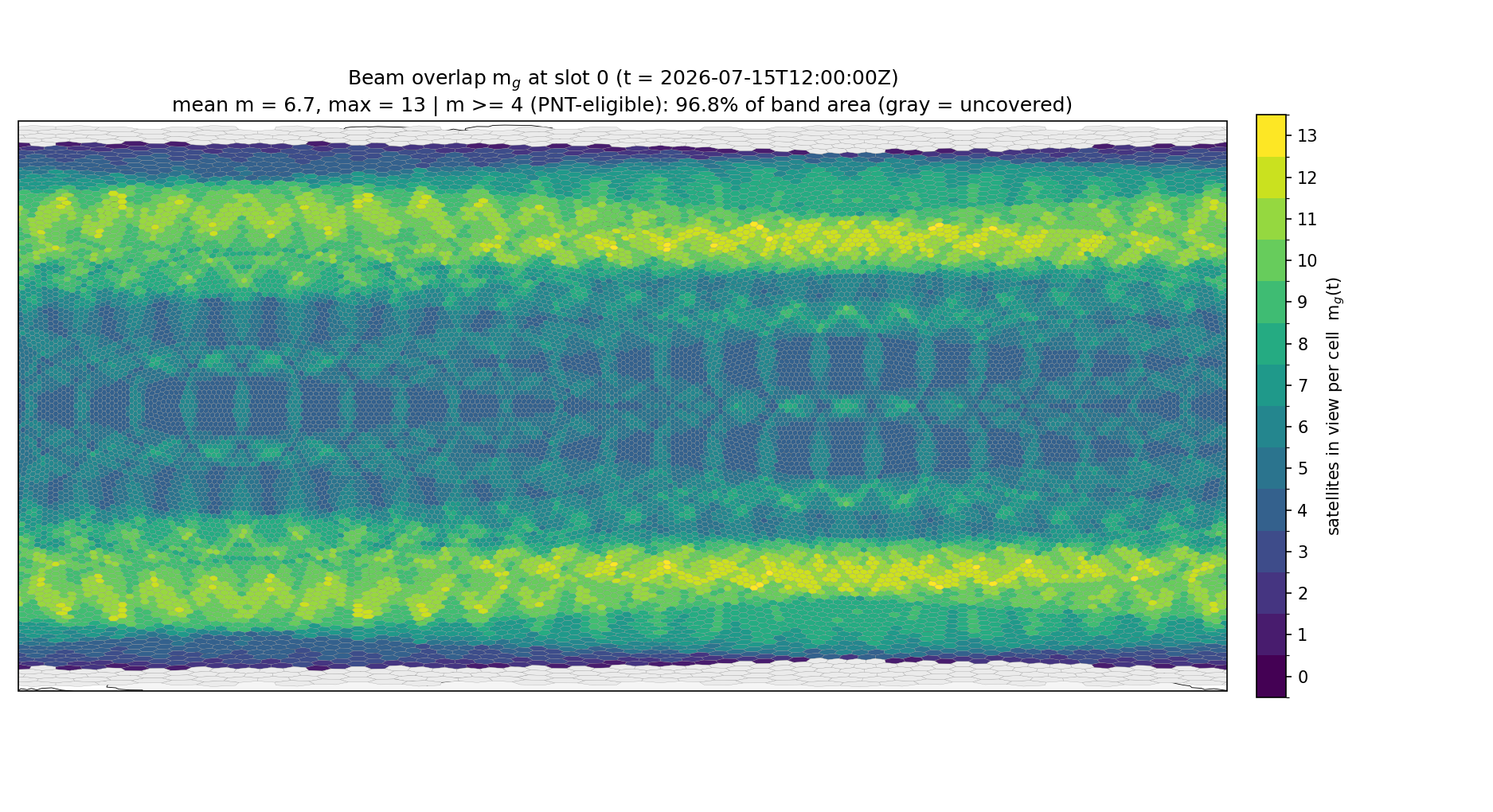}
    \caption{Overlap multiplicity $m_g$ on the ground cells at slot $t=0$.}
    \label{fig:beams_overlap}
\end{figure}

The reference simulation spans $24$~hours ($K = 1440$ slots of $60$~s, $\approx 13.2$ orbital periods) of the Walker $24\times11$ constellation at $i = 55^\circ$, $h = 1200$~km. The grid band reaches $|\phi| \le 76.7^\circ$ ($97.8\%$ of the Earth surface, $M = 21{,}380$ cells), while the actual service reach is $i + \psi = 73.2^\circ$. Figure~\ref{fig:beams_overlap} shows the overlap multiplicity $m_g$ at one time slot: the instantaneous \ac{PNT}-eligible share ($m_g \ge K_{\min}$) is remarkably stable at $94.7\%$ of the total area covered (min $94.7$, max $94.8$ over the $24$~hours). The overlap range from $m_g = 4$ near the equator to $m_g = 14$ at the $\pm50^\circ$ bands. 

\begin{figure}
    \centering
    \includegraphics[width=\linewidth]{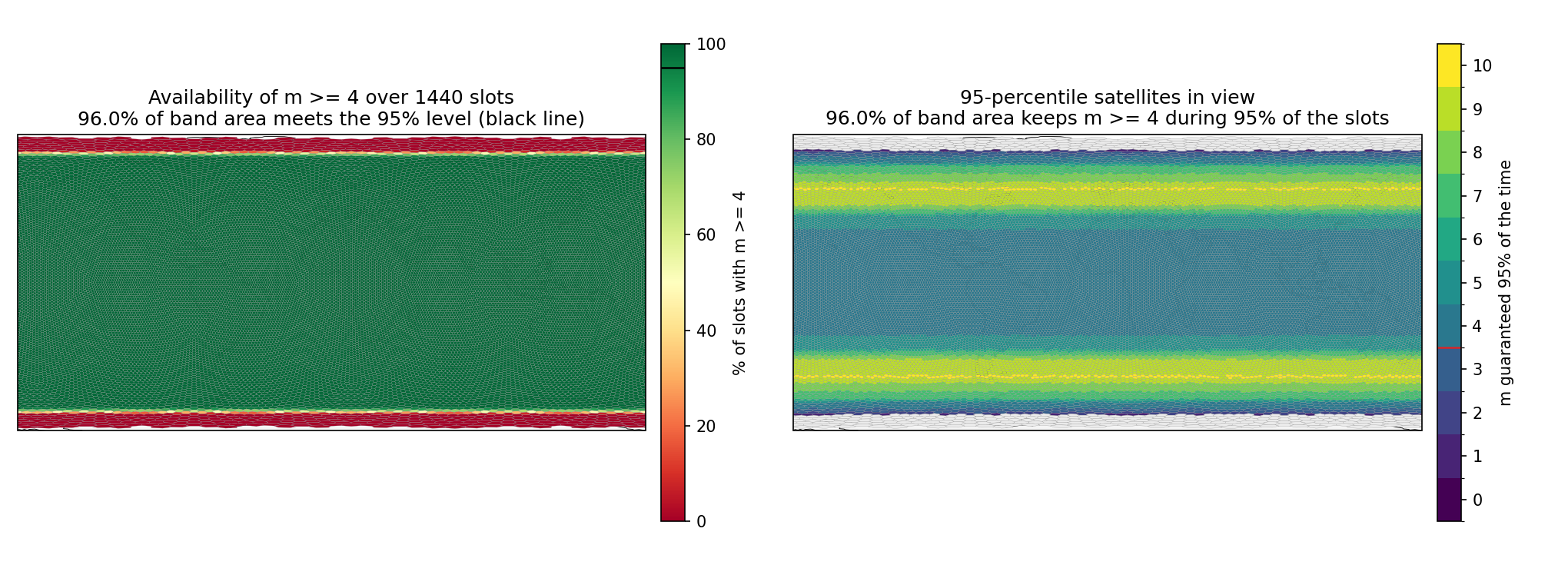}\\[2pt]
    \includegraphics[width=\linewidth]{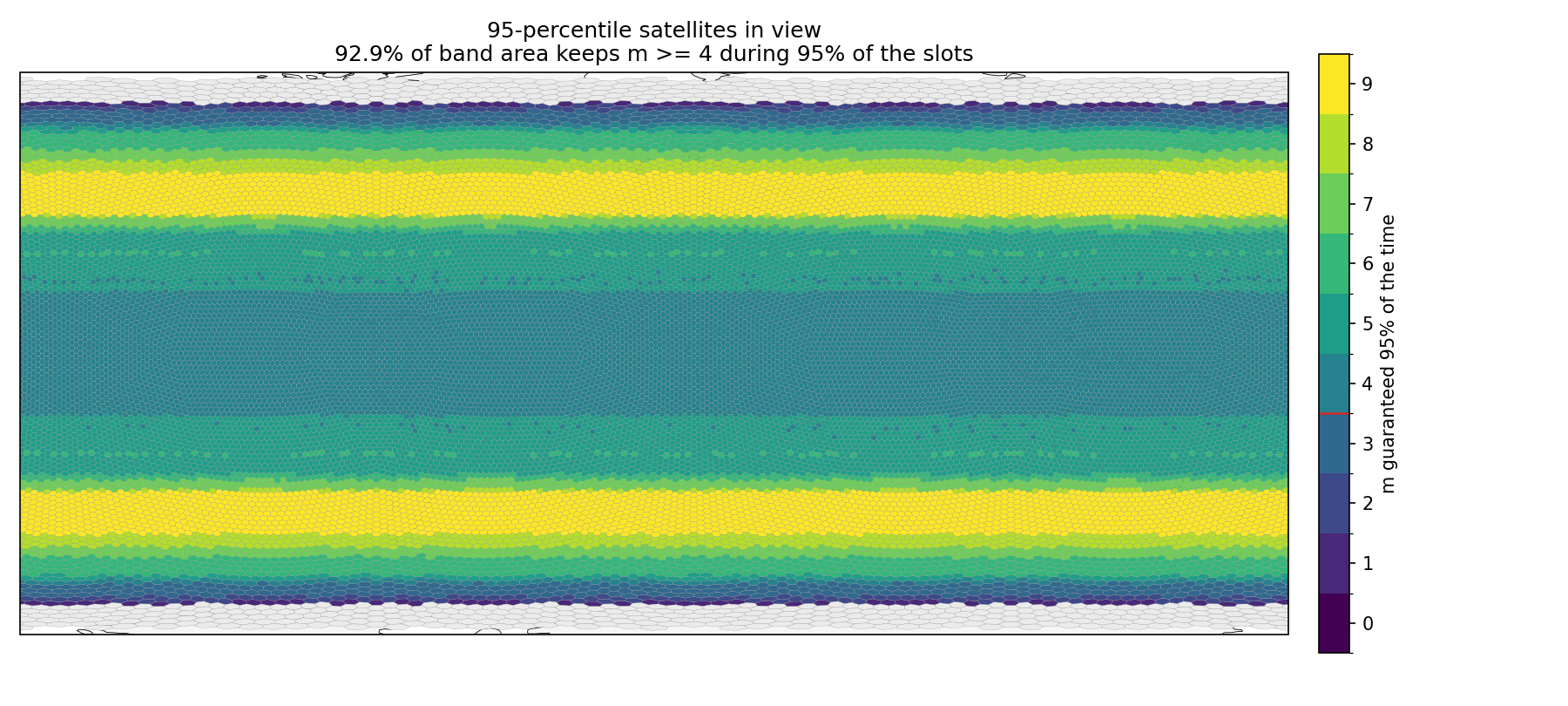}
    \caption{Eligibility availability of the campaign. Top: share of slots with $m_g \ge K_{\min}$ per cell, with the $\gamma = 95\%$ level marked on the colorbar, the cells at or above it form the geometric ceiling. Bottom: the satellite count guaranteed during $95\%$ of the slots; cells at or above the $K_{\min}$ mark hold the eligibility persistently.}
    \label{fig:availability_scan}
\end{figure}

The 95\% percentile availability yields the geometric ceiling of the scenario (see Fig.~\ref{fig:availability_scan}), $\eta_{\mathrm{geo}}^{0.95} = 92.9\%$ of the coverage area ($19{,}865$ cells). The missing $7.1\%$ are for two reasons: the never-eligible polar area ($830$ cells); and the band-edge cells whose eligibility gaps exceed the $5\%$ of no \ac{PNT} availability of more than $72$ slots.

On the traffic side, the homogeneous profile keeps $33.0\%$ of the cells busy on average (median demand $471$ of $1440$ slots per cell), whereas the population-weighted profile concentrates the same total load on $10.2\%$ of the cells (median demand $4$ slots), with the per-satellite \ac{COM} load statistics of Section~\ref{sec:power_diff} (mean $26.1$ / $8.2$, peaks $60$ / $88$ beams).

\subsection{PNT availability}

The operating point of the system is that every serviceable demanded cell is served, so the results reduce to the \ac{PNT} availability limited by the beams that are providing \ac{COM}. At $\gamma = 0.95$ only $72$ of $1440$ slots can miss the \ac{PNT} service. With no power differentiation between \ac{COM} and \ac{PNT}, compliance to the 95\% availability collapses to $\eta_{\mathrm{pnt}}^{0.95} = 0.0\%$ for homogeneous profile and $\eta_{\mathrm{pnt}}^{0.95} = 74.1\%$ for population-weighted, where only the cells that are in the ocean comply as they are almost empty (denying the availability requirement precisely where users are).

\begin{itemize}
    \item Under \emph{co-satellite sharing} (first solution, $X_{\mathrm{hot}} = 20$, $X_{\mathrm{free}} = 5$/$6$), the serve-all policy is power-feasible now at every satellite and slot and attains its eligibility ceiling exactly:
    \begin{equation}
      \eta_{\mathrm{pnt}}^{0.95}\big|_{\mathrm{(b)}} = 70.5\% \ \text{(hom.)}, \qquad 88.2\% \ \text{(pop.)},
    \end{equation}
    the limitation in both profiles to reach the geometric ceiling is that the cells with \ac{COM} demand need $m_g \ge K_{\min}+1$.

    \item Under \emph{in-beam ranging} (second solution) that requirement disappears and the ceiling is met on both profiles, $\eta_{\mathrm{pnt}}^{0.95}\big|_{\mathrm{(a)}} = 92.9\%$.
\end{itemize}

\subsection{Trade-off between the two solutions}\label{sec:tradeoff}

The strengths and weaknesses are complementary for each solution:

\begin{enumerate}
    \item Co-satellite sharing deploys with standard beams and signals: no waveform redesign, no payload coherence requirement beyond the constellation timing, receiver processing identical to a dedicated ranging service; its weaknesses are the $m_g \ge K_{\min} + 1$ eligibility penalty on busy cells, the two-tier power discipline it must enforce, and an availability guarantee that stops $22.4$~pp (homogeneous) or $4.7$~pp (population) short of the geometric ceiling.

    \item In-beam ranging reaches the ceiling and even relaxes the power problem (one ranging beam fewer per busy cell), but demands the deepest change: ranging embedded in the \ac{COM} waveform, phase coherence between the two signal components on board, and a \ac{PNT} receiver able to extract pseudoranges from a communication signal, an \ac{JCAP}-class redesign of both ends of the link~\cite{gonzalez-garrido_joint_2025, gonzalez-garrido_5g_2026}.
\end{enumerate}
 
 Under the realistic population-weighted traffic the quantitative difference is small ($88.2\%$ vs $92.9\%$): an operator can deploy the first solution with today's payloads and reserve the waveform redesign for a second generation, whereas under uniformly loaded cells the gap widens to $22.4\%$ and the second solution becomes the difference between a regional and a near-global availability guarantee. 
 
 Both solutions serve the complete \ac{COM} demand, both operate with the same two-tier power discipline, and neither requires a beam scheduler at the nominal design point.

\begin{figure*}[t]
  \centering
  \includegraphics[width=0.49\textwidth]{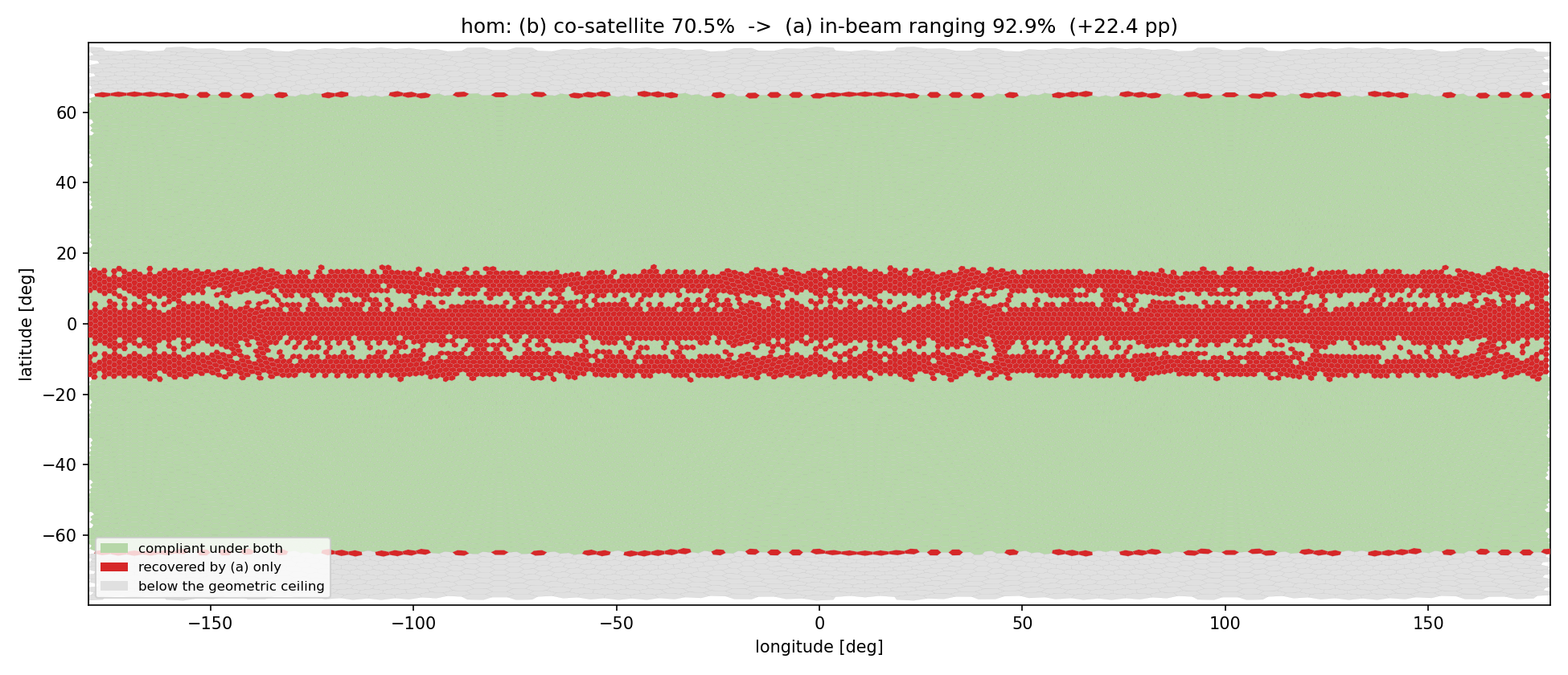}\hfill
  \includegraphics[width=0.49\textwidth]{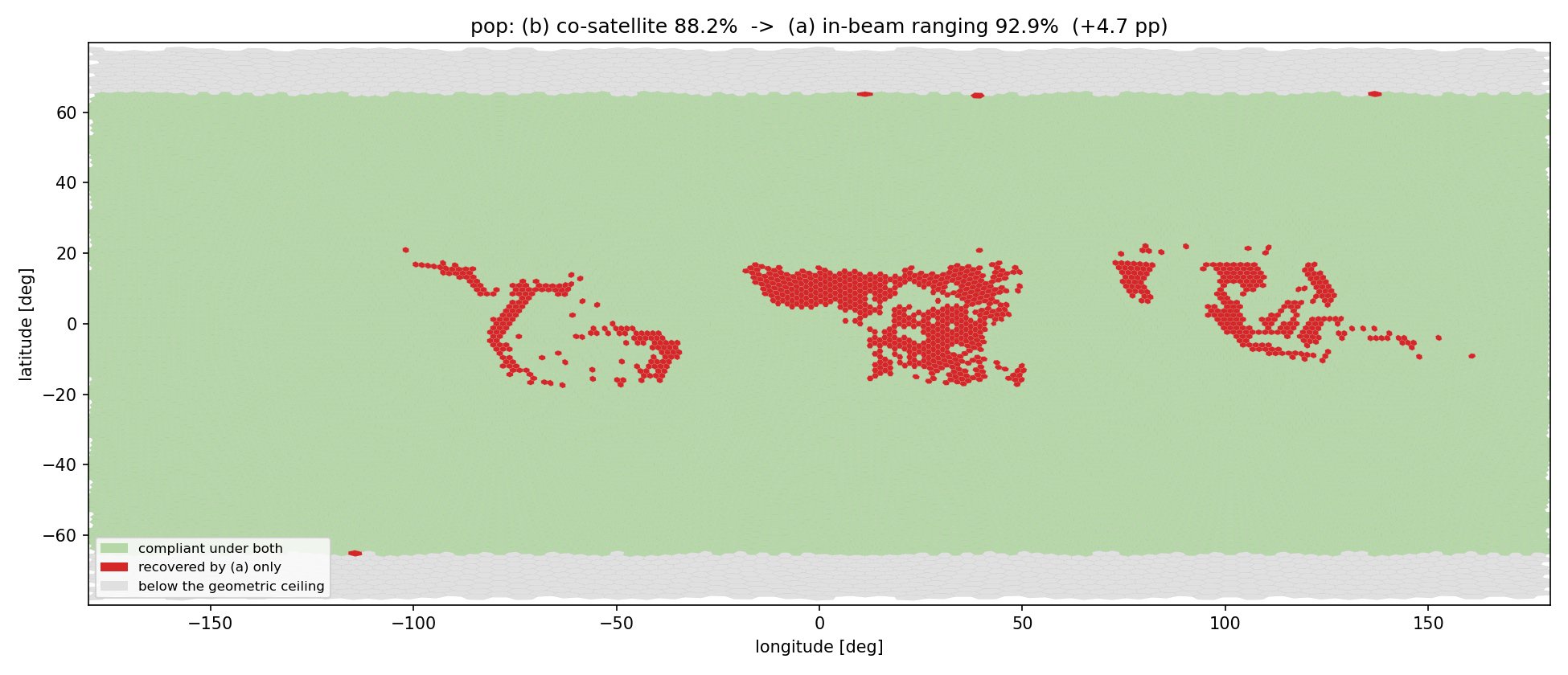}
  \caption{What in-beam ranging buys, per traffic profile: cells compliant under both solutions (green), cells recovered only by the second solution (red, concentrated in the thin-overlap equatorial belt), and cells below the geometric ceiling (gray, unreachable by any allocation). Homogeneous profile on the left, population-weighted on the right.}
  \label{fig:tradeoff}
\end{figure*}

Table~\ref{tab:tradeoff} confronts the two solutions and Fig.~\ref{fig:tradeoff} maps where the second one earns its gap. The geography is the striking part: the cells recovered by in-beam ranging ($4{,}789$ homogeneous, $1{,}015$ population-weighted) concentrate in the thin-overlap equatorial belt (median $|{\rm lat}| \approx 9^\circ$), because that is where $m_g = K_{\min}$ exactly and any busy slot breaks the $m_g \ge K_{\min}+1$ requirement of the first solution. Per service region (Fig.~\ref{fig:cells_regions}) the largest gains are Africa ($+40.1\%$ homogeneous, $+36.6\%$ population-weighted), followed by the Americas and Asia, the second solution pays off precisely over the populated equatorial land masses, while the first already closes the mid-latitude bands where the overlap is generous.

\begin{table}[t]
    \centering
    \caption{The two solutions at the COM-priority operating point (all serviceable demand served, serve-all \ac{PNT}, two-tier power with $X_{\mathrm{hot}} = 20$). $24$-hour reference campaign, $\gamma = 0.95$; geometric ceiling $92.9\%$.}
    \label{tab:tradeoff}
    \small
    \begin{tabularx}{\columnwidth}{@{}p{0.38\columnwidth} X X@{}}
        \toprule
        & (b) co-satellite & (a) in-beam ranging \\
        \midrule
        Signal / payload change
          & none & ranging embedded in the \ac{COM} waveform \\
        Busy-cell eligibility
          & $m_g \ge K_{\min}+1$ & $m_g \ge K_{\min}$ \\
        $\eta_{\mathrm{pnt}}^{0.95}$, homogeneous
          & $70.5\%$ & $92.9\%$ (= ceiling) \\
        $\eta_{\mathrm{pnt}}^{0.95}$, population
          & $88.2\%$ & $92.9\%$ (= ceiling) \\
        Ranging beams on a busy cell
          & $K_{\min}$ at $-13$~dB & $K_{\min}-1$ at $-13$~dB \\
        Quiet-beam power (serve-all feasible)
          & $-7.0$ / $-7.8$~dB & same or louder \\
        \ac{COM} demand served
          & $100\%$ & $100\%$ \\
        Scheduler required at nominal load
          & no & no \\
        \bottomrule
    \end{tabularx}
\end{table}

\section{Conclusions}\label{sec:conclusions}

This paper addressed how a broadband \ac{LEO} constellation should share its multibeam payload between communications and  \ac{PNT} services. The answer proposed here is a pair of \emph{scheduler policies}, closed form at the operating point of practical interest (communications with full priority).

Three structural results support the policies. First, the $95\%$-availability metric is governed by two bounds, a geometric ceiling fixed by the constellation ($92.9\%$ of the band area for the reference Walker $24\times11$ at $i=55^\circ$) and a beam-budget bound at the $(1-\gamma)$ rest (from the $95\%$-availability metric) rotation stretches by $1/\gamma$. The proposed power-differentiating scheme for the ranging beams moves the binding mechanism from the scheduler to the link budget. Besides, the per-satellite analysis yields the full-footprint power ratio $X^{*}(n)$ as a function of the \ac{COM} load, and could lead that an unloaded satellite illuminates its whole footprint at $-6$~dB per \ac{PNT} beam. The worst measured megacity (cells) load requires $88$ concurrent \ac{COM} beams, and the full footprint is covered reducing by $-9$~dB the \ac{PNT} beams.

The two policies proposed then differ only in the coexistence rule on a \ac{COM}-served cell. Co-satellite sharing, \ac{PNT} from the other visible satellites at a large power backoff, requires no signal modification and provides a 95\% availability area of $70.5\%$ / $88.2\%$ for homogeneous / population-weighted traffic profiles, its shortfall is confined to the thin-overlap equatorial belt. In-beam ranging, the \ac{COM} waveform carrying the \ac{PNT} signal itself, reaches the full geometric ceiling of 95\% availability ($92.9\%$) at the cost of an \ac{JCAP}-class waveform and receiver redesign, with Africa as the largest regional beneficiary (up to $+40\%$).

Future work follows the frontier. Beyond $n_{\max}(X_{\max})$, traffic growth, tighter power floors, stricter $K_{\min}$, or regional priorities under scarcity, the serve-all \ac{COM} regime ends and the scheduling problem reappears; the deficit-based availability policy and the regional priority weights developed alongside this work are the natural instruments there. On the signal side, the in-beam ranging solution calls for the waveform and receiver co-design quantified in the \ac{PRS}-coexistence literature, and the aided-fix alternative ($K_{\min} = 3$), which simultaneously cheapens every cell and raises the geometric ceiling to $95.4\%$, deserves a system-level study of its own.

\appendices
\section*{Appendix. Acronyms}
This paper uses an extensive number of acronyms, and to assist the reader, the following list presents all of them:
\begin{acronym}[TC-OFDM]
  \acro{3G}{third generation}
  \acro{4G}{fourth generation}
  \acro{5G}{fifth generation}
  \acro{6G}{sixth generation}
  \acro{3GPP}{3rd generation partnership project}
  \acro{AI}{artificial intelligence}
  \acro{AMF}{access and mobility function}
  \acro{AoA}{angle of arrival}
  \acro{AoD}{angle of departure}
  \acro{AR}{augmented reality}
  \acro{ASIC}{application-specific integrated circuit}
  \acro{ASN.1}{abstract syntax notation one}
  \acro{AWGN}{additive white Gaussian noise}
  \acro{BER}{bit error rate}
  \acro{BFN}{beamforming network}
  \acro{BO}{back-off}
  \acro{BW}{bandwidth}
  \acro{C/A}{civilian acquisition}
  \acro{CA}{cell average}
  \acro{CCDF}{complementary cumulative density function}
  \acro{CFAR}{constant false alarm rate}
  \acro{CFO}{carrier frequency offset}
  \acro{CMMB}{China mobile multimedia broadcasting}
  \acro{CP}{cyclix prefix}
  \acro{COM}{communications service}
  \acro{CRLB}{Cramer-Rao lower bound}
  \acro{CSS}{chirp spread spectrum}
  \acro{CU}{centralized unit}
  \acro{DAC}{digital to analog converter}
  \acro{DBF}{digital beamforming}
  \acro{DFT}{discrete Fourier transform}
  \acro{DL-AoD}{downlink angle of departure}
  \acro{DLL}{delay locked loop}
  \acro{DL-OTDoA}{downlink observed time difference of arrival}
  \acro{DMRS}{demodulation reference signal}
  \acro{DSSS}{direct-sequence spread spectrum}
  \acro{DU}{distributed unit}
  \acro{DVB}{digital video broadcasting}
  \acro{E-CID}{enhanced cell id}
  \acro{EO}{Earth observation}
  \acro{ECDF}{empirical cumulative density function}
  \acro{E-SMLC}{evolved serving mobile location center}
  \acro{EIRP}{equivalent isotropic radiated power}
  \acro{ECEF}{Earth-centred, Earth-fixed}
  \acro{EKF}{extended Kalman filter}
  \acro{eNB}{evolved nodeb}
  \acro{EVM}{error vector magnitude}
  \acro{FHSS}{frequency hopping spread spectrum}
  \acro{FLL}{frequency locked loop}
  \acro{FFT}{fast Fourier transform}
  \acro{FFR}{full frequency re-use}
  \acro{FoV}{field of view}
  \acro{FSPL}{free space path loss}
  \acro{FR}{frequency region}
  \acro{FR1}{frequency region 1}
  \acro{FRF3}{frequency re-use factor 3}
  \acro{GDOP}{geometrical dilution of precision}
  \acro{GEO}{geostationary Earth orbit}
  \acro{GPS}{global positioning system}
  \acro{gNB}{next generation base station}
  \acro{GNSS}{global navigation satellite system}
  \acro{GHSL}{global human settlement layer}
  \acro{GS}{ground station}
  \acro{HAPS}{high-altitude platform systems}
  \acro{HIL}{hardware-in-the-loop}
  \acro{HPA}{high power amplifier}
  \acro{IBO}{input back-off}
  \acro{ISEA3H}{icosahedral Snyder equal area aperture~3 hexagonal grid}
  \acro{ICAL}{integrated communications and localization}
  \acro{ICAN}{integrated communication and navigation}
  \acro{IDFT}{inverse discrete Fourier transform}
  \acro{IFFT}{inverse fast Fourier transform}
  \acro{IoT}{internet of things}
  \acro{IIoT}{industrial internet of things}
  \acro{ICI}{inter-carrier interference}
  \acro{IMU}{inertial measurement unit}
  \acro{ISI}{inter-symbol interference}
  \acro{ISAC}{integrated sensing and communication}
  \acro{ITU}{International Telecomunications Union}
  \acro{IMT}{International Mobile Telecommunications}
  \acro{JCAP}{joint communication and positioning}
  \acro{KPI}{key performance indicator}
  \acro{LBA}{link budget analysis}
  \acro{LCS}{location-based services}
  \acro{LEO}{low earth orbit}
  \acro{LMC}{location management component}
  \acro{LMF}{location management function}
  \acro{LMMSE}{linear minimum mean squared error}
  \acro{LNA}{low noise amplifier}
  \acro{LOS}{line of sight}
  \acro{LPP}{localization positioning protocol}
  \acro{LPPa}{localization positioning protocol annex}
  \acro{LTE}{long term evolution}
  \acro{MCRB}{modified Cramer Rao bound}
  \acro{MEO}{medium Earth orbit}
  \acro{ML}{machine learning}
  \acro{Multi-RTT}{multi-cell round trip time}
  \acro{NF}{network function}
  \acro{NSGA-II}{non-dominated sorting genetic algorithm II}
  \acro{NGSO}{non geostationary satellite orbit}
  \acro{NLOS}{non-line of sight}
  \acro{NMSE}{normalized mean square error}
  \acro{NOMA}{non-orthogonal multiple access}
  \acro{NR}{new radio}
  \acro{NTN}{non-terrestrial network}
  \acro{OFDM}{orthogonal frequency-division multiplexing}
  \acro{OMP}{orthogonal matching pursuit}
  \acro{OTA}{over-the-air}
  \acro{OTDoA}{observed time differential of arrival}
  \acro{OTFS}{orthogonal time frequency space}
  \acro{PAPR}{peak-to-average power ratio}
  \acro{PDSCH}{physical downlink shared channel}
  \acro{PLL}{phase-locked loop}
  \acro{PNT}{positioning, navigation, and timing}
  \acro{POD}{precise orbit determination}
  \acro{PPP}{precise point positioning}
  \acro{PRN}{pseudo-random noise}
  \acro{PRS}{positioning reference signal}
  \acro{PSD}{power spectral density}
  \acro{PSS}{primary synchronization signal}
  \acro{PVT}{position, velocity and timing}
  \acro{QoS}{quality of service}
  \acro{RAN}{radio access network}
  \acro{RAT}{radio-access-technology}
  \acro{RB}{resource block}
  \acro{RE}{resource element}
  \acro{RG}{resource grid}
  \acro{RedCap}{reduced capacity}
  \acro{RMSE}{root mean square error}
  \acro{ROC}{receiver operating characteristic}
  \acro{RTK}{real time kinematics}
  \acro{RRC}{radio resource control}
  \acro{SBAS}{satellite based augmentation system}
  \acro{SDR}{software defined radio}
  \acro{SIB}{system information block}
  \acro{SIC}{sequential interference cancellation}
  \acro{SIR}{signal-to-interference ratio}
  \acro{SINR}{signal-to-interference plus noise ratio}
  \acro{SFN}{single frequency network}
  \acro{SLA}{service level agreement}
  \acro{SNR}{signal-to-noise ratio}
  \acro{SoO}{signal of opportunity}
  \acro{SoP}{signal of opportunity}
  \acro{SRS}{sounding reference signal}
  \acro{SRRC}{square root raised cosine}
  \acro{SSB}{synchronization signal block}
  \acro{SSP}{subsatellite point}
  \acro{SGP4}{simplified general perturbations 4}
  \acro{SSS}{secondary synchronization signal}
  \acro{TA}{timing advance}
  \acro{TC}{time coded}
  \acro{TC-OFDM}{time-coded orthogonal frequency division multiplexing}
  \acro{TDL}{tapped delay line}
  \acro{TN}{terrestrial network}
  \acro{ToA}{time of arrival}
  \acro{ToF}{time of flight}
  \acro{TS}{technical specification}
  \acro{TR}{technical report}
  \acro{UAV}{unmanned aerial vehicle}
  \acro{UE}{user equipment}
  \acro{UL-AoA}{uplink angle of arrival}
  \acro{UL-TDoA}{uplink time difference of arrival}
  \acro{UPA}{uniform planar array}
  \acro{VR}{virtual reality}
  \acro{WLAN}{wireless local area network}
  \acro{WSG84}{world geodetic system 1984}
  \acro{ZOH}{zero-order hold}
\end{acronym}

\section*{Appendix. Notation}

\begin{table}[h]
    \centering
    \caption{Main notation of the system model.}
    \label{tab:notation}
    \begin{tabular}{l l}
        \toprule
        Symbol & Meaning \\
        \midrule
        $R$ &  Earth radius \\
        $\mathcal{G}$, $M$ & cell set and its cardinality \\
        $A_g$, $\mathbf{u}_g$ & area and centroid unit vector of cell $g$ \\
        $w$, $D$ & cell width and beam diameter across flats \\
        $\mathcal{S}$, $S$ & satellite set, $S = PQ$ \\
        $h$, $i$, $F$ & altitude, inclination, Walker phasing \\
        $N_{\mathrm{b}}$, $N_{\max}$ & beams per satellite, active beam budget \\
        $\mathcal{F}_s(t)$, $\psi$ & footprint of satellite $s$, cap radius \\
        $v_{s,g}(t)$ & visibility indicator \\
        $\mathcal{V}_g(t)$, $m_g(t)$ & visiting set of cell $g$ and its size \\
        $s^{\star}(g,t)$, $\rho_{s,g}(t)$ & serving satellite, slant range \\
        $K$, $\Delta t$ & number of slots, slot duration \\
        $K_{\min}$ & minimum satellites for a \ac{PNT} fix \\
        $\lambda$, $\bar{L}$, $a$ & arrival rate, mean session length, load \\
        $\delta_g(t)$ & demand indicator of cell $g$ \\
        $c_g(t)$, $\pi_g(t)$ & \ac{COM} and \ac{PNT} service indicators \\
        $J_{\mathrm{COM}}$, $\eta_{\mathrm{com}}$ & \ac{COM} service metrics in \eqref{eq:objectives} \\
        $\alpha_g$, $\eta_{\mathrm{pnt}}^{\gamma}$, $w_g$ & \ac{PNT} availability, metric, region weight \\
        $X$; $X_{\mathrm{free}}$, $X_{\mathrm{hot}}$ & COM-to-\ac{PNT} per-beam power ratios (two-tier) \\
        $b_g$ & per-slot beam cost of cell $g$ in \eqref{eq:budget_rotation} \\
        \bottomrule
    \end{tabular}
\end{table}

\section*{Acknowledgement}
During the preparation of this work the author(s) used Claude/Fable and Google/Gemini 3.6 in order to proofread the text and validate/cross-check the mathematical derivations and simulation scripts. After using these tools, the author(s) reviewed and edited the content as needed and take(s) full responsibility for the content of the publication.

\bibliographystyle{IEEEtran}
\bibliography{references}

% \begin{IEEEbiography}
% [{\includegraphics[width=1in,height=1.25in,clip,keepaspectratio]{Alejandro.jpg}}]
% {Alejandro Gonzalez-Garrido} ~ PhD student at the SIGCOM group of SnT (University of Luxembourg), specializing in hybrid GNSS and 5G \ac{PNT} systems using Non-Terrestrial Networks. Holds an integrated degree and an M.Sc. in Telecommunication Engineering, obtained in 2015. Has professional experience in the timing and synchronization industry, satellite design, and network operations.
% \end{IEEEbiography}

\end{document}